\documentclass{emulateapj}

\begin{document}

\title{High mass star formation in normal late-type galaxies: observational constraints 
   to the IMF}
\author{
A. Boselli\altaffilmark{1}, S. Boissier\altaffilmark{1}, L. Cortese\altaffilmark{2}, V. Buat\altaffilmark{1}, T.M. Hughes\altaffilmark{2} 
G. Gavazzi\altaffilmark{3}
}
\altaffiltext{1}{Laboratoire d'Astrophysique de Marseille, UMR 6110 CNRS, 38 rue F. Joliot-Curie, F-13388 Marseille France}
\altaffiltext{2}{School of Physics and Astronomy, Cardiff University, 5, The Parade, Cardiff CF24 3YB, UK}
\altaffiltext{3}{Universita degli Studi di Milano-Bicocca, Piazza delle Scienze 3, 20126 Milano, Italy}

\begin{abstract}
We use H$\alpha$ and far-ultraviolet (FUV, 1539 \AA) GALEX data for a large sample of nearby objects to study the high mass ($m$ $\geq$ 2 M$\odot$) star formation 
activity of normal late-type galaxies. The data are corrected for dust attenuation using the most accurate techniques at present available,
namely the Balmer decrement for H$\alpha$ data and the total far-infrared to FUV flux ratio for GALEX data. The sample shows a 
highly dispersed distribution in the H$\alpha$ to FUV flux ratio (Log $f(H\alpha)/f(FUV)$ = 1.10 $\pm$ 0.34 \AA) indicating that two of the most
commonly used star formation tracers give star formation rates with uncertainties up to a factor of 2-3. The high dispersion is partly due to the presence of AGN,
where the UV and the H$\alpha$ emission can be contaminated by nuclear activity,
highly inclined galaxies, for which the applied extinction corrections are probably inaccurate, 
or starburst galaxies, where the stationarity in the star formation history required for transforming H$\alpha$ and UV 
luminosities into star formation rates is not satisfied. Excluding these objects, normal late-type galaxies have 
Log $f(H\alpha)/f(FUV)$ = 0.94 $\pm$ 0.16 \AA, which corresponds to an uncertainty of $\sim$ 50\% on the SFR.
The H$\alpha$ to FUV flux ratio of the observed galaxies increases with their total stellar mass. If limited to 
normal star forming galaxies, however, this relationship reduces to a weak trend that might be totally removed using different
extinction correction recipes. In these objects the H$\alpha$ to FUV flux ratio seems also barely related with the FUV-H colour, 
the H band effective surface brightness, the total star formation activity and the gas fraction. 
The data are consistent with a Kroupa (2001) and Salpeter initial mass function in the high mass stellar range ($m$ $>$ 2 M$\odot$) and imply,
for a Salpeter IMF, that the variations of the slope $\gamma$ cannot exceed 0.25, from $\gamma$ = 2.35 for massive 
galaxies to $\gamma$ = 2.60 in low luminosity systems. We show however that these observed trends, if real, can 
be due to the different micro history of star formation in massive galaxies with respect to dwarf systems. 
\end{abstract}


\setcounter{footnote}{0} 

\section{Introduction}

The formation and evolution of galaxies can be
observationally constrained through the study of their present and
past star formation activity. While the star formation history of
galaxies can be well determined by fitting their stellar ultraviolet (UV) to near-infrared
spectral energy distribution with population synthesis models, the
direct observation of the youngest high mass stars is generally used
to infer their ongoing activity. The present day star
formation rate (SFR) is generally determined using the luminosity of the
youngest stars and the relation $SFR$ = $K(\lambda)$ $L(\lambda)$,
where $K(\lambda)$ can be inferred from population synthesis models
under several assumptions (see Kennicutt 1998 for a review). Hydrogen Balmer emission lines and UV fluxes, both
related with the emission of the youngest stellar populations,
are thus often used as direct tracers of the present day star formation activity of galaxies.
This widely used technique can be blindly applied only if the initial mass function (IMF) is
universal, thus independent of morphological type, luminosity and
redshift and if the star formation activity of the target galaxies has
been constant for a time $\geq$ than the lifetime of the emitting
stars, i.e. $\sim$ 10$^7$ years for Balmer emission lines and a few
10$^8$ years for the UV ($\lambda$ $<$ 2000 \AA). \\
Direct measurements of the IMF in the Milky
Way or in other very nearby galaxies (Massey et al. 1995; Scalo 1998; Selman \& Melnick
2008) are consistent with a universal IMF (e.g. Scalo 1986; Kroupa 2001; Renzini
2005). A few recent indirect observational results on the mass to light ratio of 
low surface brightness galaxies (Lee et al. 2004), on the SED properties 
of SDSS (Hoversten \& Galzebrook 2008) and starburst (Rieke et al. 1993) galaxies, 
on the H$\alpha$ to UV flux ratio of nearby
objects (Meurer et al. 2009) as well as on the star formation activity
vs. stellar mass assembly in the far universe (Dav\'e 2008; Wilkins et
al. 2008), combined with theoretical (Krumholz \&
McKee 2008) and statistical considerations
(Pflamm-Altenburg et al. 2007) led several authors to question this
statement. These works suggest that the IMF, in particular when
considered as a unique function representative of the whole galaxy,
might not be universal but rather changing with the galaxy stellar
mass, the star formation activity or the gas column density. The IMF is a key
parameter in the process of star formation: proving its non
universality would thus have enormous implications on the study of the
star formation process at all scales, from the physics of the
interstellar medium to the formation and evolution of galaxies since
the earliest phases of the universe. Indeed the IMF is a key
ingredient in all models of galaxy evolution. 
\\
It should also be tested whether the condition of stationarity in star formation, required 
for transforming H$\alpha$ and UV fluxes into star formation rates, 
is satisfied in all kinds of galaxies. There are indeed some indications that this is not the case: 
it has been shown that a bursty
history of star formation is expected in dwarf galaxies (Fioc \& Rocca-Volmerange 1999),
consistent with the determination of the star formation history obtained using the 
stellar colour-magnitude diagram of dwarf galaxies in the local group (e.g. Mateo 1998).
A recent starburst activity has been also proposed to explain the 
the H$\alpha$ to UV flux ratio of UV luminous galaxies (Sullivan et al. 2000).\\
It is thus time to investigate whether these two assumptions on the
constancy of the star formation history of late-type galaxies and on
the universality of the IMF are consistent with the latest
observational data in our hands. The comparison of the star formation activity determined
using independent indicators of a well defined sample of galaxies with
a complete dataset of H$\alpha$, UV and IR imaging and optical ($R$
$\sim$ 1000) integrated spectroscopy allows us to quantify the
uncertainty associated to different tracers. Furthermore since the
ratio of the number of ionizing and non-ionizing photons emitted by a
galaxy depends on the slope and on the upper mass cutoff of the IMF,
on its present and past star formation activity and on its
metallicity, the H$\alpha$ to UV flux ratio (properly corrected for
dust attenuation)
can be used to constrain the shape of the IMF.\\
Determining the UV ionizing and non-ionizing emitted radiation through
observations is however extremely difficult since at these wavelengths
the emission of galaxies is severely attenuated by the dust of the
interstellar medium. Both UV continuum and Balmer line data need thus
to be corrected for dust attenuation.  Quantifying the dust
attenuation of the UV radiation is still one of the major challenges
in modern astronomy. Models and observations have recently shown that
dust extinction can be well determined considering that the UV
radiation absorbed by dust is re-emitted in the far-IR (Buat \& Xu
1996; Witt \& Gordon 2000; Buat et al. 2002; Boselli et al. 2003; Cortese et al. 2008).  At
the same time the attenuation of the Balmer lines can be determined
using the Balmer decrement (e.g. Lequeux et al. 1981). This is
generally done by comparing the observed to the nominal H$\alpha$ to H$\beta$ flux ratio, this last being
fairly constant with a value of 2.86 (case B, Osterbrock 1989),
provided that spectroscopic data have a sufficient resolution for
deblending the [NII] from H$\alpha$ and for accurately measuring the
underlying Balmer absorption. The ionizing radiation could also be
absorbed by dust or escape the galaxy before ionizing the gas. All
these phenomena should be taken into account in the determination of
the emitted radiation using observational data.\\
The study of the shape of the IMF using H$\alpha$ and UV data was 
first proposed by Buat et al. (1987) and later adopted by
Meurer et al. (1999), Boselli et al. (2001) and Charlot et al. (2002).
Although limited by the statistics of their sample and by the lack of
corollary data for accurate extinction corrections, all these works
agreed with a constant IMF. The advent of the GALEX mission
(Martin et al. 2005), which provided us with homogeneous UV data for
thousands of objects, and the availability of high quality integrated
spectroscopy for hundreds of galaxies in the nearby universe
(Kennicutt 1992; Jansen et al. 2000; Gavazzi et al. 2004; Moustakas \&
Kennicutt 2006) with IR data, are providing us with a unique opportunity to revisit this subject
with a new dataset. The work of Salim et al. (2007), based on GALEX
data and SDSS spectra, or that of Meurer et al. (2009) suggested 
that the H$\alpha$ to UV flux ratio could indeed
vary with stellar mass.  \\
The aim of the present work is to study the high mass star
formation activity of normal, late-type galaxies. The
paper is structured as follows: in sect. 2 we give a brief description
of the sample, in sect. 3 of the dataset. The analysis and the discussion are
given in sect. 4 and 5 respectively. A detailed description of the
observational data (A), of the derived parameters (B) used in the analysis
with their uncertainties (C) and on possible selection biases (D) are
presented in the Appendix.

\section{The sample}

The analysis presented in this work is based on three samples of late-type nearby galaxies with available H$\alpha$ and UV data: 
the Herschel Reference Survey (Boselli et al. 2009), which is a K band selected ($K_{Stot}$ $\leq$ 12 mag), 
volume limited (15$<$ $Dist$ $<$ 25 Mpc) sample of galaxies,
the Virgo cluster sample extracted from the Virgo
Cluster Catalogue (VCC) of Binggeli et al. (1985), an optically selected sample complete to $m_B$ $\leq$18 mag ($M_B$ $\leq$ -13), 
and the UV Atlas of galaxies of Gil de Paz et al. (2007).\\
To exclude those objects whose interaction with the cluster environment might have perturbed and
modified their star formation history, we removed all galaxies with an HI-deficiency $>$ 0.4 (Boselli \& Gavazzi 2006).
The interaction with the cluster environment might indeed change the H$\alpha$ over UV flux ratio (Iglesias-Paramo et al. 2004) as in the 
case of the anemic galaxy NGC 4569 (Boselli et al. 2006).
For the same reason we also excluded strongly interacting or merging galaxies belonging to the GALEX UV atlas.\\
The resulting sample, not complete in any sense, includes thus 198 galaxies with H$\alpha$ and FUV data (see Table \ref{Tabcomp}) 
and spans a large range in morphological type (from Sa to Sm, Im and BCD)
and luminosity (-22 $<$ $M_B$ $<$ -13). Dwarf systems (-15$<$ $M_B$ $<$ -13) are present only
in the Virgo cluster sample.\\
Distances are determined assuming galaxies in Hubble flow with H$_0$ = 75 km s$^{-1}$ Mpc$^{-1}$
except for Virgo cluster objects where distances are taken according to Gavazzi et al. (1999).\\

\section{The data}

The study of the high mass star formation activity of nearby,
late-type galaxies is based on the analysis of their H$\alpha$
and UV emission. All sample galaxies thus have narrow band H$\alpha$
and broad band GALEX imaging data. Multifrequency imaging (optical,
near-IR and far-IR) and spectroscopy (optical, HI and CO) data are
also necessary to derive the different entities used in the following
analysis, such as stellar masses, metallicities, effective surface
brightnesses, colours, gas column densities and deficiencies (see the Appendix B).\\
H$\alpha$ imaging data are corrected for [NII] contamination and dust
extinction using long slit, integrated spectroscopy with a resolution
$R$ $\sim$ 1000, optimal for deblending [NII] from H$\alpha$ and for
resolving the underlying H$\beta$ absorption. If
spectroscopic data are not available, then H$\alpha$ imaging data are
corrected for [NII] contamination and dust attenuation using standard,
luminosity dependent statistical relationships (see the Appendix B).  UV
GALEX data are corrected for dust extinction using the prescriptions
of Cortese et al. (2008) based on TIR/FUV flux ratio whenever IRAS
far-IR data are available, or using standard recipes based
on the UV spectral slope or H band effective surface brightness in other cases (Cortese et al. 2006).\\
The analyzed sample can thus be divided into three different
subsamples according to the quality of the available data (see Table
\ref{Tabsamp}). The {\it high quality sample} (H.Q.) includes 111 galaxies
for which the Balmer decrement and the TIR/FUV flux ratio guarantee
the best possible dust extinction correction.
The {\it medium quality sample} (M.Q.) is formed by 148
galaxies\footnote{H.Q $\subset$ M.Q. $\subset$ L.Q.} includes also 
objects where dust extinction has been corrected
using the Balmer decrement for the H$\alpha$ imaging data and
indirect, empirical relationships for the FUV fluxes whenever TIR data
are not available.
The {\it low quality sample} (L.Q.) includes in addition
galaxies for which both H$\alpha$ and UV data have been corrected
using statistical and indirect relations. This last sample
includes 198 objects.
Details on the adopted dust extinction corrections
and on the different derived parameters used in the following analysis
(metallicities, stellar masses and star formation rates) are given in the
Appendix B. The Appendix (C) also gives an estimate of the errors on the
different variables, with a discussion on possible biases in the data (D).



  \begin{figure}
   \centering
   \includegraphics[width=15cm,angle=0]{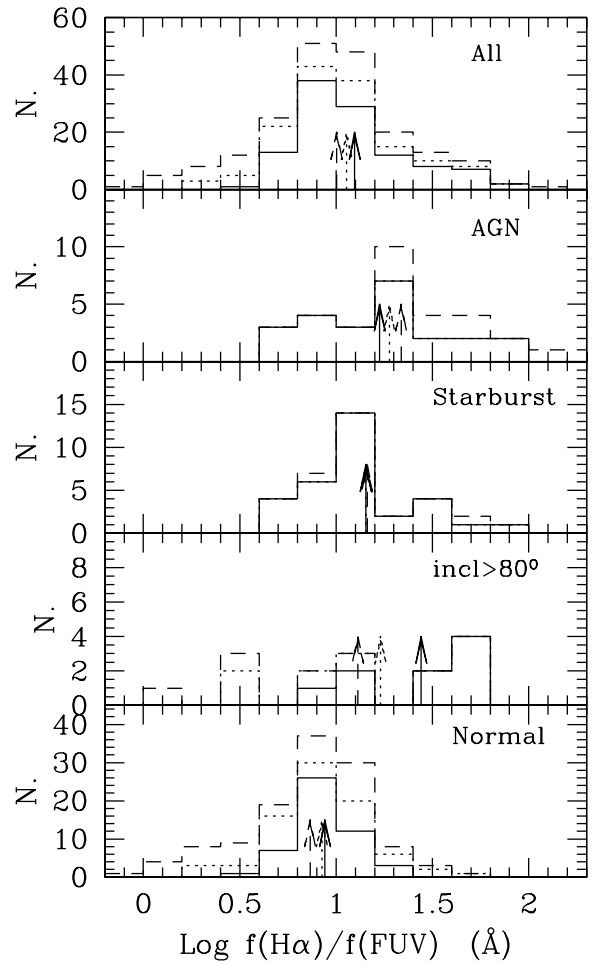}
   \caption{The H$\alpha$ to FUV flux ratio distribution (in logarithmic scale)
   for galaxies with high quality (solid line), medium quality (dotted line)
   and all (dashed line) dust extinction corrections. From top to bottom, all sample galaxies, AGNs, starbursts,
   galaxies with an inclination $>$ 80 $^o$ and normal galaxies used in the analysis of the IMF. 
   The vertical arrows indicate
   the mean values of the distributions (see Table \ref{Tabratio}).
    }
   \label{distrall}
  \end{figure}

\section{Analysis}

The H$\alpha$ luminosity gives a measure of the global
photoionization rate of the interstellar medium due to high mass ($m$
$>$ 10 ${\rm M\odot}$), young ($\le$ 10$^7$ years) O-B stars
(Kennicutt 1998). Assuming that the SFR of galaxies is constant on a
time scale comparable to the life time of the stellar population
responsible for the ionizing emission (a few 10$^7$ years), the SFR 
is proportional to the H$\alpha$ luminosity.
The UV emission of a galaxy at 1500-2300 \AA~ is dominated by the
emission of less recent ($\sim$ 10$^8$ years) and massive (2$<$ $m$
$<$ 5 ${\rm M\odot}$) A stars (Lequeux 1988). The UV emission becomes
stationary if the SFR is constant over $\sim$ 10$^8$ years in FUV 
(e.g., Boissier et al., 2008). In that case
the relatively high mass star formation activity of a galaxy is
proportional to the FUV luminosity. Whenever the star
formation activity is constant over $\sim$ 10$^8$ years, the H$\alpha$
to FUV flux ratio is proportional to the ratio of the number of
ionizing to non-ionizing massive stars and is thus directly related to
the shape of the IMF in the high mass star range ($m$ $>$ 2 M$\odot$).
To simplify the analysis, we assume that the fraction of ionizing
photons that escape the galaxies before ionizing the gas (escape
fraction) is zero. We also assume that the fraction
of ionizing photons absorbed by dust before ionizing the gas is zero ($f$=1). \\
The top panel of Fig. \ref{distrall} shows the distribution of the
H$\alpha$ over FUV flux ratio for the whole sample (dashed line). The first
interesting result is that the dispersion in the distribution of the
selected galaxies is quite large 
even when we limit this estimate to the high quality sample (0.34 dex, solid line, see Table \ref{Tabratio}), for which
uncertainties due to dust extinction corrections are less severe.
Since this large dispersion might be due to the presence of AGNs
(where part of the global H$\alpha$ and UV emission might not be
related to star formation and where the [NII] contamination to the
narrow band imaging data is probably more uncertain than in normal
galaxies), starbursts (where star formation is known to vary on short
time scales) or highly inclined galaxies (where the extinction
correction are highly uncertain), we separate the sample in four
different subsamples: AGNs (Seyfert 1 and 2, LINER), starbursts
(defined as those galaxies with a 60 over 100 $\mu$m IRAS flux ratio
$F(60)/F(100)>0.5$, Rowan-Robinson \& Crawford 1989), edge-on galaxies (with an inclination $>$ 80$^o$,
but including also the other highly inclined objects listed in Table \ref{Tabestremi}) and
the remaining star forming galaxies ($F(60)/F(100)$$\leq$0.5,
inclination $\leq$ 80$^o$, non AGN), that hereafter we call ``normal''
to differentiate them from those belonging to the previous categories.
The dispersion of the H$\alpha$ over FUV flux ratio of normal, star
forming galaxies (0.16 dex) is, as expected, significantly reduced
with respect to the original sample. Despite their small number,
highly inclined galaxies have the highest H$\alpha$ to FUV flux ratio
(Log $f(H\alpha)/f(FUV)$ = 1.44 $\pm$ 0.32 \AA). Most of them have
prominent dust lanes, as indicated in Table \ref{Tabestremi}. This
result is in agreement with Panuzzo et al. (2003) and Tuffs et al. (2004) who used dust models
for different geometries. They claimed that standard recipes for dust
extinction correction calibrated on face-on or low inclination
galaxies underestimate $A(FUV)$ in edge-on systems.  AGNs, and in a
minor way, starbursts also have H$\alpha$ to FUV flux ratios larger
than normal galaxies (see Table \ref{Tabratio}).
A Kolmogorov-Smirnov test has shown that the probability that the distribution of Log $f(H\alpha)$/$f(FUV)$ 
for AGN, starbursts and inclined galaxies is similar to that of "normal" 
galaxies is 1\%, 34\% and 2\% respectively. 
\\

\subsection{The H$\alpha$ to FUV vs. $M_{star}$ relation}

\begin{figure}
  \centering
  \includegraphics[width=15cm,angle=0]{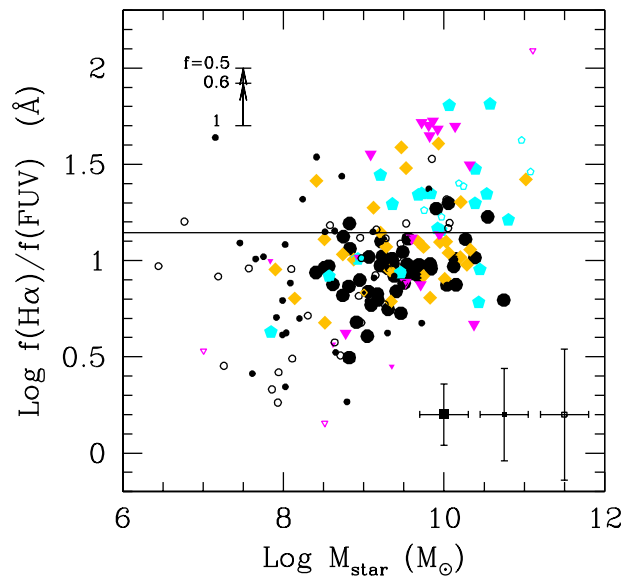}
  \caption{The relationship between the H$\alpha$ to FUV flux ratio
    and the total stellar mass (both in logarithmic scales) for all
    galaxies in the sample, in units of erg cm$^{-2}$ s$^{-1}$
    (H$\alpha$) and erg cm$^{-2}$ s$^{-1}$ \AA$^{-1}$ (FUV).  Large
    filled symbols are for the high quality sample, 
    small filled symbols for the medium quality sample and small empty symbols for the low quality sample. 
    Magenta triangles are for highly inclined ($incl$ $>$ 80$^o$)
    galaxies, including those objects listed in Table
    \ref{Tabestremi}, orange diamonds for starburst ($F60/F100$ $>$ 0.5)
    galaxies, cyan pentagons for AGNs, while black symbols for ``normal''
    galaxies. The solid line gives the expected ratio for the standard
    Kennicutt (1998) values. The
    vertical black arrows give the Y-axis shifts expected for different
    corrections considering the fraction of ionizing photons absorbed
    by dust ($f$ =1 is for no correction). The error bars give the
    typical uncertainties for galaxies in the high, medium and low
    quality samples.  }
  \label{hafuvmstar}
\end{figure}

Several observational
evidences indicate that the galaxy mass is the most important parameter
driving galaxy evolution (Cowie et al. 1996; Gavazzi et al. 1996;
Boselli et al. 2001; Gavazzi et al. 2002b). Figure \ref{hafuvmstar}
shows a relationship between Log $f(H\alpha)/f(FUV)$ and the galaxy
stellar mass (compared to the constant ratio determined for a Salpeter\footnote{Although never directly
measured down to the hydrogen burning limit, a Salpeter IMF is used as reference
throughtout the paper since generally used in extragalactic studies.} IMF using
the calibration of Kennicutt 1998, black solid line), with massive galaxies having on average a higher ratio
than dwarfs (Table \ref{Tabmass}) with however considerable scatter.
This relation is still present when only high quality data are used (big
filled symbols), but it is significantly reduced (the dynamic range in Log
$f(H\alpha)/f(FUV)$ drops from $\sim$ 1.4 to $\sim$ 0.3; see below) if highly inclined
(magenta triangles), AGN (cyan pentagons) and starburst galaxies
(orange diamonds) are not considered. 

Using SDSS spectroscopy and UV GALEX data of a large sample of star
forming galaxies Salim et al. (2007) found a similar trend between the
H$\alpha$ over FUV ratio and the total stellar mass of galaxies. They  
imputed their result as due to an inaccurate dust
extinction correction based on standard Charlot \& Fall (2000) models
calibrated on the UV slope.
The trend observed in our sample ($\sim$ 1.4 dex) is more important
than that of Salim et al. (2007) which is $\sim$ 0.5 dex within a mass
range 10$^9$ $\leq$ $M_{star}$ $\leq$ 10$^{11}$ M$\odot$ probably
because of the larger dynamic range in stellar
mass of our sample (10$^7$ $\leq$ $M_{star}$ $\leq$ 10$^{11}$ M$\odot$) 
and the presence of AGN, excluded in Salim et al. (2007).
Finally, a trend between the H$\alpha$ to UV flux ratio and the R band
luminosity or the rotational velocity of the galaxies has been found
by Meurer et al. (2009), with an amplitude ($\sim$ 1 in dex)
comparable to ours. A similar
dependence of the H$\alpha$ to FUV flux ratio with the B band absolute magnitude or the star formation rate
has been shown by Lee et al. (2009) on a nearby sample dominated by dwarf galaxies.
While Salim et al. (2007) imputed the observed trend to an inaccurate dust extinction correction,
Meurer et al. (2009)  interpreted it as a clear evidence of a varying IMF with galaxy mass.
Lee et al. (2009) concluded that, although the data are consistent with a changing IMF,
other causes can be at the origin of the observed trend.

\section{Discussion}

The observed variation of the H$\alpha$ over FUV flux ratio as a function of the galaxy stellar 
mass might have different origins: it might be due to an inaccurate  
dust attenuation correction, to a variation of the
escape fraction and dust absorption of ionizing photons,
to metallicity effects, to the lack of stationarity for the star formation activity 
over the last few million years, or to a variation of the IMF. With the aim of reducing as much as possible
the degrees of freedom in the adopted dataset, we exclude from the following analysis all those objects
where the determination of the corrected H$\alpha$ and FUV fluxes can be complicated by the presence
of active galactic nuclei (AGN), extremely inclined galaxies ($incl$ $>$ 80$^o$), where the extinction
correction is known to be problematic, and starburst galaxies, where the star formation activity is known to vary
on short time scales. As previously shown (Fig. \ref{distrall}) these galaxies have an H$\alpha$ to FUV flux ratio
significantly different than that of normal, star forming objects. 
We thus focus our attention to the ``normal'' galaxy sample, for which a trend between the H$\alpha$ to FUV flux ratio
and the stellar mass seems still present (black symbols in Fig. \ref{hafuvmstar}).

\begin{figure*}
  \centering
  \includegraphics[width=15cm,angle=0]{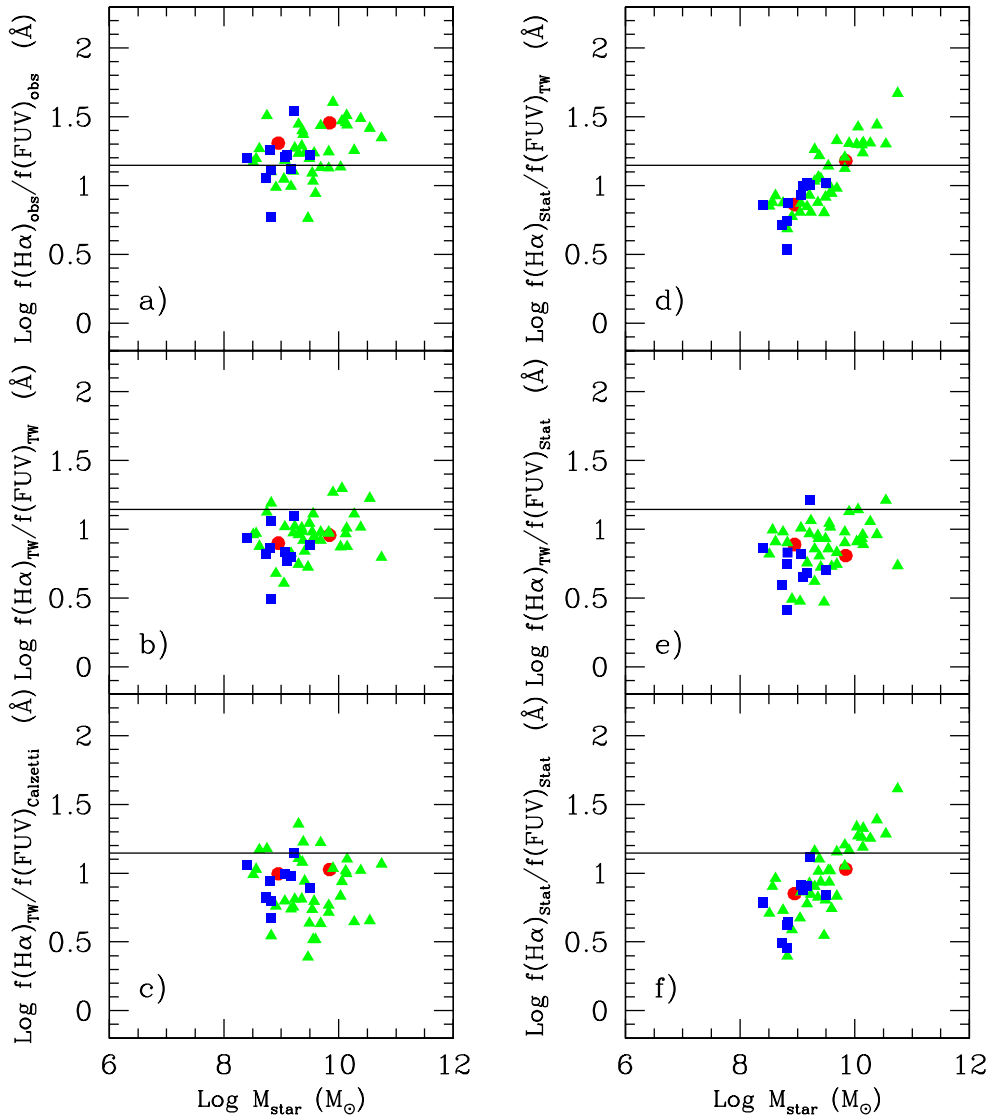}
  \caption{The relationship between the H$\alpha$ to FUV flux ratio
    and the stellar mass (both in logarithmic scales) for ``normal''
    galaxies belonging to the high quality sample, where the suffix "obs" stands 
    for observed entities, "TW" for the standard corrections applied in this work, 
    "Calzetti" for corrections based on the Calzetti's law and "Stat"
    for statistical corrections similar to those used by Meurer et al. (2009) (see text).
    The solid,
    horizontal line gives the expected ratio for the Kennicutt (1998)
    calibrations. Red circles are for Sa-Sb, green triangles for Sbc-Sd and
    blue squares for Sm-Im-BCD galaxies. The best fit to the data are given in Table \ref{Tabcor}. }
  \label{hafuvmstarcomp}
\end{figure*}

\subsection{Dust extinction}

Extinction corrections for both the FUV and H$\alpha$ data are still the
highest source of uncertainty in the determination of the emitted
H$\alpha$ over FUV flux ratio even in normal galaxies.  We should thus
be sure that the observed trends shown in the previous figures do not
result from any systematic effect in the applied corrections.  
To see whether extinction corrections can be at the origin of the
observed trend in Fig. \ref{hafuvmstar}, we plot in Fig.
\ref{hafuvmstarcomp} the relationship between the Log of the H$\alpha$
to FUV flux ratio and the Log of the stellar mass using data corrected
following different dust attenuation recipes. To reduce the
uncertainties, we limit this comparison to the high quality sample.
The plotted relationships are for: a) observed entities, b) H$\alpha$
and FUV data corrected for dust attenuation as described in the
present work (Balmer decrement for H$\alpha$ and the TIR/FUV
calibration of Cortese et al. 2008 for the FUV), c) H$\alpha$ data
corrected using the Balmer decrement and FUV data corrected following
the prescription for starburst galaxies of Calzetti (2001): $A(FUV)$ =
1.78 $A(H\alpha)$, d) H$\alpha$ data corrected using the luminosity
dependent statistical relation given in the Appendix B and FUV data
corrected as described in this work, e) H$\alpha$ data corrected using
the Balmer decrement and FUV data corrected using the TIR/FUV vs.
FUV-NUV relation given by Mu\~noz-Mateos et al. (2009) obtained for
normal, late-type galaxies,
$L_{TIR}/L_{FUV}=10^{0.30+1.15(FUV-NUV)}-1.64$, combined with the
$A(FUV)$ vs. $L_{TIR}/L_{FUV}$ relationship given by Buat et al.
(2005), and finally f) H$\alpha$ data corrected using the luminosity
dependent statistical relation given in the Appendix B and FUV data using
the FUV-NUV colour index as in e). The corrections used in f) are
analogues to those adopted by Meurer et al. (2009), where $A(H\alpha$)
is determined from an $A(H\alpha$) vs. R band luminosity relation and
$A(FUV)$ using a $L_{TIR}/L_{FUV}$ vs. FUV-NUV
relation. The corrections in d) and e) are analogues to the Meurer et al. (2009) method for
$A(H\alpha$) and $A(FUV)$ respectively.\\
Figure \ref{hafuvmstarcomp} and Table \ref{Tabcor} show that a trend in the H$\alpha$ to FUV
flux ratio vs. $M_{star}$ relationship is present, although with a
large scatter and a small dynamic range ($\sim$ 0.6 dex) when
uncorrected data are used (a). The steepness of the relationship is
reduced when data are corrected using our recipes (b), and are totally
removed when FUV data are corrected using the Calzetti's law (c).
Opposite to this, a very tight relationship is observed when a
luminosity dependent correction for the H$\alpha$ data is adopted (d),
while a dispersed trend similar to that present in the observed data
is seen when the FUV data are corrected using a recipe based on the
FUV-NUV colour (e). The combination of a luminosity dependent
correction for H$\alpha$ and a FUV-NUV colour dependent correction for
FUV data introduces a strong and tight correlation between the
H$\alpha$ to FUV flux ratio
and the galaxy stellar mass (f) (see Table \ref{Tabcor}).\\
This comparison indicates that adopting the 
statistical, luminosity dependent extinction correction given in the
Appendix B and shown in Fig. \ref{niiha} for the H$\alpha$ data is unsuited since it
introduces systematic effects in the data. This is quite obvious given that the H
band luminosity (from which $A(H\alpha)$ is determined) is strongly 
correlated to the galaxy stellar mass. The Balmer decrement is thus
the most critical parameter for determining the emitted H$\alpha$ to
FUV flux ratio.  The low quality sample is thus totally unreliable for
any statistical study and will thus not be used in the following
analysis. Corrections of the FUV data based on the FUV-NUV
colour\footnote{These are the corrections generally  used in the medium quality sample} (e) do
not introduce any systematic effect since here they conserve the same trend
present in the observed data. For this reason the medium quality
sample could still be used, although with caution, in the analysis.
Once the Balmer decrement is available, the adoption of different
recipes for correcting the FUV data can maintain (e), reduce (b) or
remove (c) the observed trend between Log $f(H\alpha)/f(FUV)$ and Log
$M_{star}$. This means that any observed variation of the H$\alpha$ to
FUV flux ratio with other galaxy parameters might result from
systematic effects introduced by the adopted correction and should
thus be considered with extreme caution.  As a first consideration we
can say that the relationships between the H$\alpha$ to FUV flux ratio
and the R band luminosity or the rotational velocity (which is tightly
related with the total mass of galaxies through the virial theorem)
shown by Meurer et al. (2009) can result from their adopted
corrections, in particular the H$\alpha$ one which is determined using
an $A(H\alpha)$ vs. $L_R$ relation. In a more indirect way, the trend
between H$\alpha$ to FUV flux ratio and the R band or H$\alpha$
surface brightness shown in Meurer et al. (2009) might also be
affected by the adopted corrections since the surface brightness of
the old and young stellar populations are also tightly related to the
total mass of galaxies (Gavazzi et al. 1996).  To conclude, for
constraining the high mass star formation activity of late-type
galaxies high quality, medium resolution integrated spectroscopy (for
determining the Balmer decrement and the H$\alpha$ to [NII] ratio)
is mandatory. To a minor extent, IR data necessary for correcting UV fluxes are also necessary. \\

\begin{figure}
  \centering
  \includegraphics[width=15cm,angle=0]{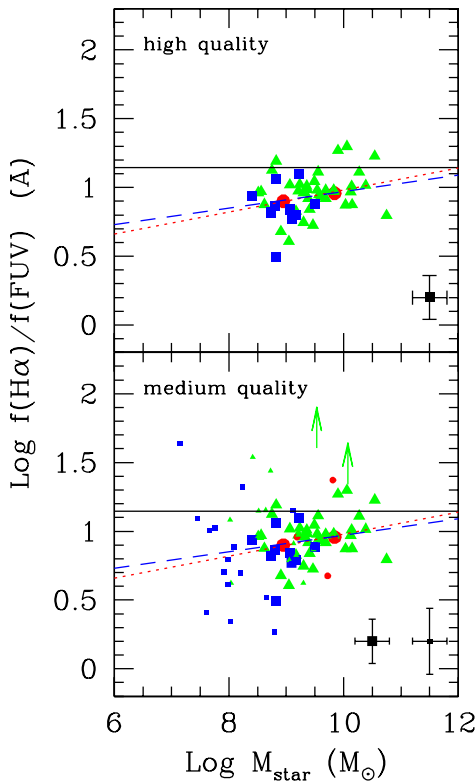}
  \caption{The relationship between the H$\alpha$ to FUV flux ratio
    and the stellar mass (both in logarithmic scales) for ``normal''
    galaxies belonging to the high quality sample (upper panel; big symbols) and to the medium quality sample (lower panel; small symbols).
    The solid, horizontal line gives the expected ratio for the Kennicutt (1998)
    calibrations, the red, dotted line the best fit to the high quality sample and the blue, dashed line to the medium quality sample (see Table \ref{Tabfit}). Red circles are for Sa-Sb, green triangles for Sbc-Sd and
    blue squares for Sm-Im-BCD galaxies.  }
  \label{hafuvmstarext}
\end{figure}

The correction recipes adopted in this work are likely the more
reliable to date since directly based on the TIR to FUV ratio of normal spirals and not on empirical relations (e) or
calibrated on starburst galaxies (c) (see Buat et al. 2002).
We remind however that even in our recipes global FUV-optical/near-infrared colours are needed
to convert the TIR/FUV ratio into A(FUV). Thus we cannot completely exclude that a weak "spurious" 
trend in colour (and perhaps in stellar mass) is introduced by the dust corrections technique here adopted.
\\
Once corrected adopting our recipes (Cortese et al. 2008), the H$\alpha$ to FUV flux ratio still shows a very weak trend
with stellar mass, with massive objects having slightly higher values of Log $f(H\alpha)/f(FUV)$
than dwarfs (0.14 dex), as shown in Table \ref{Tabmass}. The Spearman probability that the two variables are correlated 
is however weak (94\%) (\ref{Tabfit}). The dynamic range covered by Log$f(H\alpha)/f(FUV)$ is comparable to the 
statistical uncertainty on this variable (see the Appendix C). It is thus impossible to state whether the observed trend is real
or due to systematic effects related to the adopted extinction correction. A complete set of multiwavelength data for dwarf galaxies is
necessary for extending the dynamic range in stellar mass and thus proving the real existence of this relationship.
Lee et al. (2009) have recently done a similar analysis on a sample of nearby galaxies in the local universe
using a set of multifrequency data similar to ours (H$\alpha$ narrow band imaging and FUV GALEX fluxes, corrected for extinction
using integrated spectroscopy and far-IR data). 
Their analysis has shown that the H$\alpha$ to FUV flux ratio decreases in low luminosity systems (their sample includes galaxies down to
$M_B$ $\sim$ -12, while our is mostly limited to $M_B$ $\sim$ -15).
A tentative extension of our analysis to the low luminosity regime can be done by adding the medium quality sample, for which Balmer decrement measurements are 
available. As shown in Table \ref{Tabfit} and in Figure \ref{hafuvmstarext}, the relationship between the H$\alpha$ to FUV flux ratio 
and $M_{star}$ (and $(FUV-H)_{AB}$ and $\mu_{e}(H)_{AB}$, see next Sect.) extends also to the medium quality sample at the same significance level.
At the same time statistical considerations (see Sect. 5.5.1) predict that the integrated IMF of a given galaxy 
is truncated at high masses whenever its global star formation is low, as is the case in dwarf systems.
We thus consider the H$\alpha$ to FUV flux ratio vs. $M_{star}$ relation as real, and we try to understand its origin.

\subsubsection{Dust absorption of the ionizing radiation}

In the case that dust is mixed with gas, only a fraction of the Lyman continuum photons ($f$) 
produced in the star forming regions contributes to the ionization of the atomic hydrogen, the remaining
(1-$f$) being absorbed by dust (e.g. Inoue et al. 2000). The determination of $f$ from an observational point of view 
is quite problematic since it requires the knowledge of the exact number of ionizing and non-ionizing UV photons
produced by the different stellar populations inhabiting a galaxy. Here we measure $f$ by comparing the mean 
ratio of the observed H$\alpha$ to FUV flux of massive galaxies to the expected values for different population synthesis models
\footnote {Only massive galaxies are used since these are the only objects for which the required stationarity condition for star formation is guaranteed
(see Sect. 5.4.2).}. 
Hirashita et al. (2003), using a relatively small sample of star forming galaxies with both H$\alpha$ and UV (2000 \AA)
measurements, concluded that $f$ is $\sim$ 0.57 for a Salpeter IMF in the mass range 0.1$<$ $m$ $<$100 M$\odot$, 
a value that they found almost independent on metallicity
(Hirashita et al. 2001). 
The $f(H\alpha)/f(FUV)$ flux ratio of normal, massive galaxies (Log$f(H\alpha)/f(FUV)$ = 1.03 $\pm$ 0.16 \AA)
corresponds to the expected value for a Salpeter IMF using the Kennicutt (1998) calibration (Log$f(H\alpha)/f(FUV)$ = 1.15 \AA) if $f$ = 0.77.
Different values of $f$ are obtained for other IMF, as listed in Table \ref{Tabf1},
or for a given IMF in different ranges of stellar mass. The values of $f$ determined here are larger than that of Hirashita et al. (2003),
and range from $\sim$ 1 to 0.77 for a Salpeter IMF.
It is quite unlikely that the observed trend between the H$\alpha$ to FUV flux ratio and the stellar mass in normal, 
star forming galaxies (black symbols in Fig. \ref{hafuvmstar}) is due 
to a decreasing $f$ with stellar mass since, given the strong metallicity-luminosity relation,
we expect that the fraction of ionizing photons absorbed by dust is more important in massive
galaxies than in metal poor dwarfs.

\subsection{Lyman continuum escape fraction}

A fraction of the Lyman continuum photons can escape galaxies without ionizing the surrounding hydrogen.
This effect has to be considered in the determination of the emitted ionizing radiation using H$\alpha$
observed fluxes. 
Direct measurements of the escape fraction based on the observations of the Lyman continuum photons 
are available only for starburst galaxies: they all indicate that the escape fraction is $\leq$ 6 \%
(e.g. Leitherer et al. 1995 ($<$ 3\%), Heckman et al. 2001 ($\leq$ 6\%), Deharveng et al. 2001 ($\leq$ 6\%); 
Hayes et al. 2007 (3\%)).
By measuring the ionizing radiation of the Magellanic stream, Bland-Hawthorn \& Maloney (1999) have estimated
that the escape fraction of the Milky Way is $\approx 6$\%. This value is consistent with the
poorly constrained upper limit given by Zurita et al. (2000) ($\leq$ 50\%) determined by measuring the 
contribution of the diffuse gas to the ionizing radiation of spiral discs. Given these small 
values ($\leq$ 6\%), we confidently think that a mass dependent variation of the Lyman continuum escape fraction
may not entirely cause the observed trend between the H$\alpha$ to FUV flux ratio and the galaxy stellar mass
(Fig. \ref{hafuvmstar}).

\begin{figure*}
  \centering
   \includegraphics[width=15cm,angle=0]{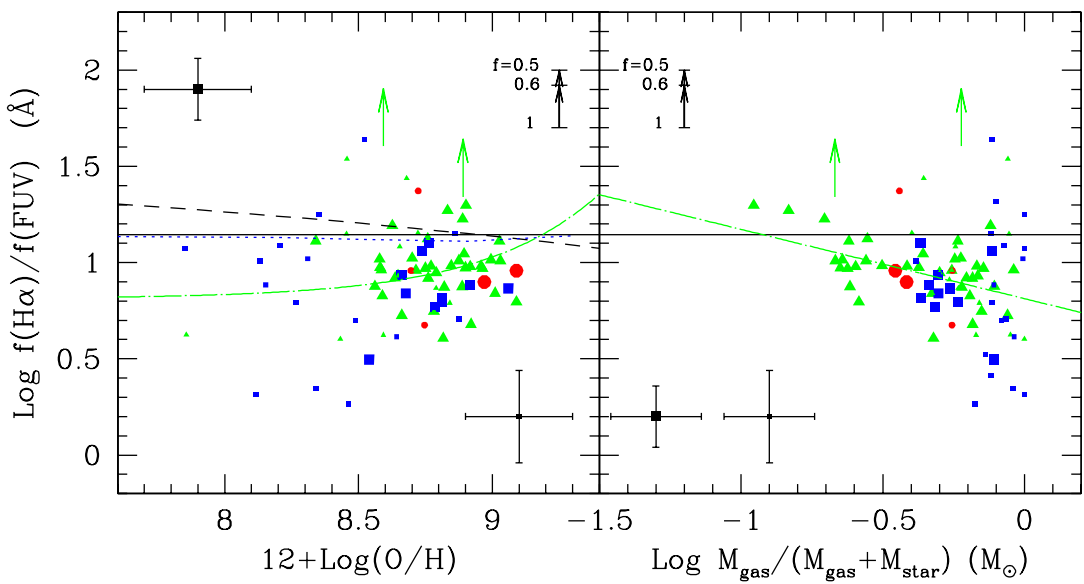}
   \caption{The relationship between the H$\alpha$ to FUV flux ratio
     and the 12 $+$ Log(O/H) metallicity index (left) and the gas mass fraction (right) 
     for normal, late-type
     galaxies in the high (large, filled symbols) and medium (small
     filled symbols) quality samples.  
     Red circles are for Sa-Sb, green triangles for Sbc-Sd
     and blue squares for Sm-Im-BCD galaxies. The error bars give the
     typical uncertainties for galaxies in the high and medium quality
     samples.
     The solid, horizontal line gives the expected ratio for the
     Kennicutt (1998) calibration, the blue dotted line for the metallicity
     dependent models of Bicker \& Fritze-Alvensleben (2005) and the
     black dashed line for the metallicity dependent models of
     Vaszquez et al. (2007) in the case that rotation of massive stars
     is taken in consideration. The green, long dashed-dotted line shows 
     the expected dependence of Log $f(H\alpha)/f(FUV)$ on 12+Log(O/H) (left panel) considering
     the observed relationship between Log $f(H\alpha)/f(FUV)$ and the 
     gas fraction (right panel; see Table \ref{Tabfit}).}
   \label{hafuvmet}
 \end{figure*}

\subsection{Metallicity}

Stellar population synthesis models indicate that the H$\alpha$ to FUV flux ratio might change with metallicity. 
Variations of $\sim$ 30 \% in H$\alpha$ and $\sim$ 20 \% in 
FUV (Bicker \& Fritze-Alvensleben 2005) are expected for metallicities between 1 and 1/50 solar.
The effects of metallicity on the $f(H\alpha)/f(FUV)$ ratio, however,
are minor since they neutralize themselves in the ratio.
In the case that the rotation of massive stars is taken in consideration (V{\'a}zquez et al. 2007), 
however, the H$\alpha$ flux can change up
to $\sim$ 30 \% while the FUV flux by only a few \% in FUV.\\
Figure \ref{hafuvmet} shows the relationship between the
$f(H\alpha)/f(FUV)$ ratio and the galaxy metallicity (12+Log(O/H)).
No relation is observed between the two variables (see Table \ref{Tabfit}). The variation of the H$\alpha$ to FUV
flux ratio predicted by the models in the range of metallicity of the sample galaxies is  
significantly smaller than the dispersion in the data. In a closed box model, the 
gas mass fraction, defined as $(M_{gas}/(M_{gas}+M_{star}))$, 
is tightly related to metallicity (Garnett 2002). The total gas content of our sample galaxies
can be determined by adding the HI measurement to the molecular hydrogen (when not available, we 
consider $M(H_2)$ = 15\% $M(HI)$, Boselli et al. 2002b) and considering 30\% of helium.
Figure \ref{hafuvmet} shows a relationship
between $f(H\alpha)/f(FUV)$ and the total gas fraction
($M_{gas}$/($M_{gas}$$+$$M_{star}$)), with lower values observed in those objects 
with higher gas fractions. 
We use the relation in Table \ref{Tabfit} (green dotted-dashed line in right panel) and
combine it with gas fraction-metallicity relation for a close box model
($Z$ = -$p$ln($M_{gas}$/($M_{gas}$$+$$M_{star})$), where $p$ is the oxygen yield; Garnett 2002),
to predict the dependence of
Log$f(H\alpha)/f(FUV)$ on 12+Log(O/H) (green, dashed-dotted line in left panel).
We find that the relatively loose trend with the gas fraction is consistent with the observed scatter in the
Log$f(H\alpha)/f(FUV)$ vs. metallicity plot.\\
Although normal galaxies follow a well known luminosity-metallicity relation 
(Zaritsky et al. 1994; Tremonti et al. 2004) 
we can exclude that the trend shown in Fig. \ref{hafuvmstarcomp} is due to metallicity since
population synthesis models predict lower values of the H$\alpha$ to FUV flux ratio for metal rich, massive galaxies,
contrary to what is observed.

  \begin{figure}
   \centering
   \includegraphics[width=15cm,angle=0]{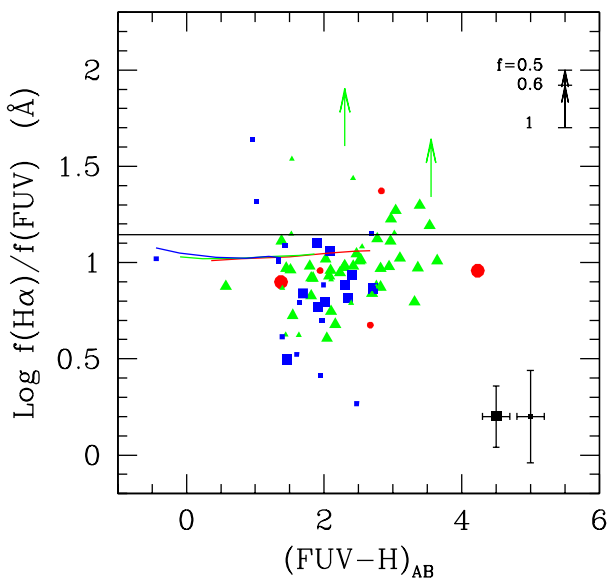}
   \caption{The relationship between the H$\alpha$ to FUV flux ratio 
   and the extinction corrected FUV-H colour index (in AB system) for normal, late-type galaxies.
   Symbols are coded as in Fig. \ref{hafuvmet}. The coloured solid lines are for our multizone,
     chemo-spectrophotometric models of galaxy evolution for galaxies with different spin parameters (red:
     $\lambda$ = 0.02, compact galaxies, green: $\lambda$ = 0.05,
     standard galaxies, blue: $\lambda$ = 0.09, low surface brightness
     galaxies).  
.
    }
   \label{hafuvfh}
  \end{figure}

  \begin{figure}
   \centering
   \includegraphics[width=15cm,angle=0]{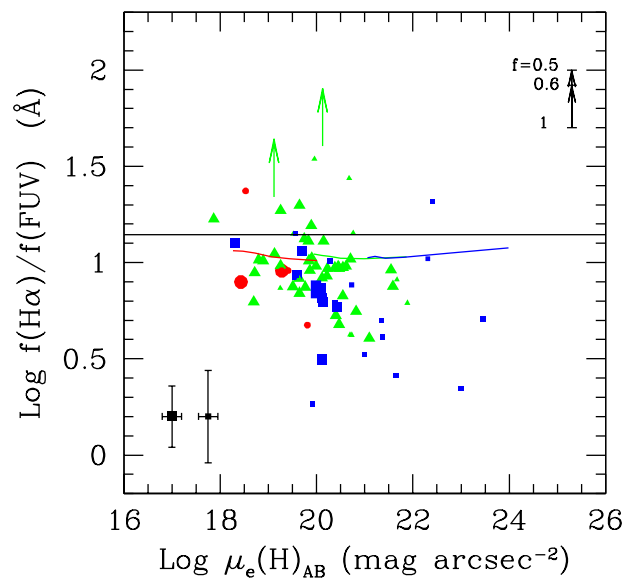}
   \caption{The relationship between the H$\alpha$ to FUV flux ratio
   and the H band effective surface brightness for normal, late-type galaxies. Symbols and lines are coded as in Fig. \ref{hafuvfh}.
    }
   \label{hafuvmuh}
  \end{figure}

\subsection{Star formation history}

\subsubsection{Secular evolution of galaxies}

Given the strong relationship between the secular evolution of the star formation history of
galaxies and their total dynamical or stellar mass (Cowie et al. 1996;
Gavazzi et al. 1996; Boselli et al. 2001; Gavazzi et al. 2002b), we
should first consider whether long term variations of the star formation activity
of the target galaxies, which span four orders of magnitude in stellar mass, 
might induce variations in the H$\alpha$ to FUV flux ratio.
To do that we plot the variation of the H$\alpha$ to FUV flux ratios as a
function of the stellar mass (not shown), FUV-H colour, H band effective surface
brightness\footnote{Defined as the surface brightness within the
  effective radius, the radius including half of the total light
  (Gavazzi et al. 2000b)} 
in Fig. \ref{hafuvfh} and
\ref{hafuvmuh} and compare the results to the predictions
of the secular chemo-spectrophotometric models of galaxy evolution of Boissier \& Prantzos (1999, 2000)\footnote{These models,
computed assuming a Kroupa (2001) IMF, are done for galaxies with rotational velocities from 40 to 360 km s$^{-1}$
and spin parameters $\lambda$ = 0.02 (red solid line, for compact galaxies, 
$\lambda$ = 0.05 (green solid line, for standard galaxies) and $\lambda$ = 0.09 (blue
solid line, for low surface brightness objects).}.
As for the stellar mass, a trend is present when the H$\alpha$ to FUV
flux ratios is plotted vs. the FUV-H colour (99 \% of probability that the two variables are 
correlated, Table \ref{Tabfit}) or the H band effective
surface brightness (99\%).
Figures
\ref{hafuvfh} and \ref{hafuvmuh} indicate that the different secular evolution of the star formation activity
of galaxies of different mass
can induce variations in the H$\alpha$ over FUV flux ratio only
up to 0.1 dex and thus cannot explain the observed trends in the
figures, where Log $f(H\alpha)/f(FUV)$ varies by 0.7 dex.

  \begin{figure}
   \centering
   \includegraphics[width=15cm,angle=0]{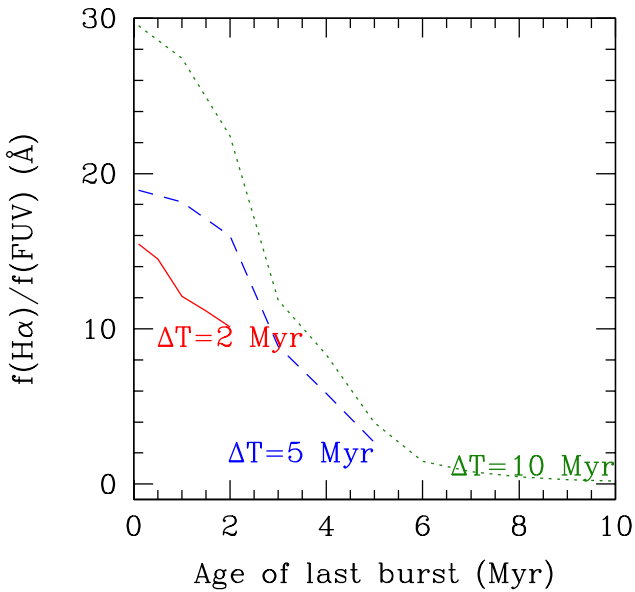}
   \caption{The variation of the H$\alpha$ over FUV flux ratio as a function of the age of the last burst as 
   expected for a galaxy whose emission is due to the contribution of 
   subsequent instantaneous bursts 
   of star formation (HII regions) every 2 (solid red), 5 (dashed blue) and 10 (dotted green) Myr.
    }
   \label{modsam1}
  \end{figure}

  \begin{figure}
   \centering
   \includegraphics[width=15cm,angle=0]{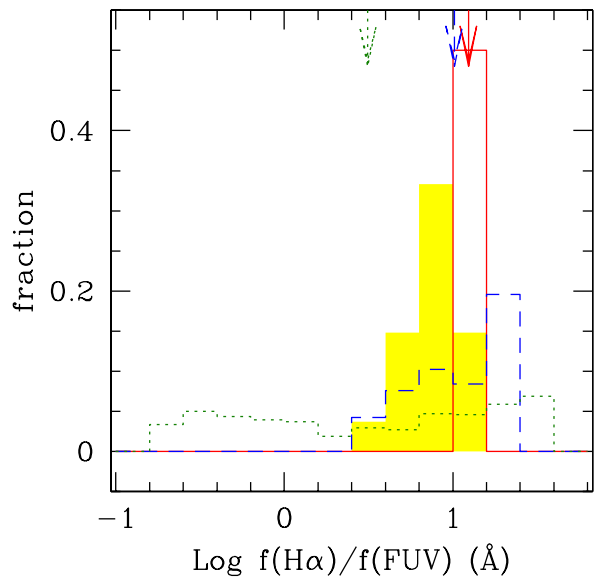}
   \caption{The expected distribution of the H$\alpha$ over FUV flux ratio for a galaxy whose emission is due to the contribution of 
   subsequent instantaneous bursts 
   of star formation (HII regions) every 2 (solid red line), 5 (dashed blue line) and 10 (dotted green line) Myr 
   compared to the observed distribution of normal, low mass ($M_{star}$ $\leq$ 10$^{9.2}$ M$\odot$)
   galaxies (yellow shaded histogram). Vertical arrows indicate the mean values. 
    }
   \label{modsam2}
  \end{figure}

\subsubsection{Micro-history of star formation}

It has been claimed that the observed variation of the
H$\alpha$ over UV ratio in star forming regions in the outer disc of
M81 is due to an age effect (Gogarten et al. 2009). 
An increase of the dispersion of the $f(H\alpha)/f(FUV)$ ratio in low
mass galaxies is expected since in these objects the H$\alpha$
emission is dominated by giant HII regions (Kennicutt 1988; Kennicutt
et al. 1989) similar to 30 Doradus in the LMC which is responsible to
$\sim$ 50\% of the galaxy emission\footnote{The contribution of 30
  Doradus to the total H$\alpha$ emission of the LMC might even be
  larger if we consider that an important fraction of the diffuse
  emission of the galaxy, which is 35\% of the total, might result
  from the ionization of the diffuse ISM due to the leakage of photons
  produced in giant HII regions.} (Kennicutt et al. 1995). Given their
short lifetime (some 10$^6$ years), comparable to that of the OB stars
responsible for the H$\alpha$ emission, with respect to that of the A
stars emitting in FUV ($\sim$ 10$^8$ yrs), the probability of seeing
a HII region still active in FUV but already quenched in H$\alpha$ is
relatively high. The observation of a galaxy whose dominant HII
region is too old to be detected in H$\alpha$ but not in FUV would
easily lower its $f(H\alpha)/f(FUV)$ ratio by a factor of two in
dwarfs, while the effect in massive spirals
would be less than a few \% just because the H$\alpha$ emission is never dominated by a single HII region.\\
Local age effects could thus contribute to the observed variation of
the H$\alpha$ to FUV flux ratio with stellar mass observed in Fig.
\ref{hafuvmstar}. There are indeed some observational evidences that
HII regions located along spiral arms are on average younger than
those observed in low density regions such as the interarm, the outer
discs of high surface brightness galaxies or those in low surface
brightness objects (von Hippel \& Bothun 1990; Oey \& Clarke 1998;
Helmboldt et al. 2008).  This evidence is probably related to the fact
that the youngest HII regions located along the spiral arms are those
newly formed during the increase of the gas column density induced by
the spiral wave. Being formed through induced star formation, they are
now undergoing a burst of activity.  Those in the interarm region were
previously formed during the last crossing of the density
wave\footnote{The timescale between the passage of spiral density
  waves is of the order of $\sim$ 40 Myr at any location along the
  disc (Rand 1993), a timescale longer than the expected life of HII
  regions (Oey \& Clarke 1998).}. Similarly, HII regions in the outer
discs of spirals, in low surface brightness Im galaxies and in BCDs
are sporadically formed thus on average older ($\sim$ 6 Myr) than
those belonging to spiral arms that can be considered as coeval
($\leq$ 1 Myr, Helmboldt et al. 2008; Tamburro et al. 2008).  As for
the observed decrease of the H$\alpha$ E.W. of HII regions
sporadically formed (von Hippel \& Bothun 1990), we could expect that
the same HII regions have a lower H$\alpha$ to FUV flux ratio with
respect to those still undergoing a starburst since formed through an
induced process (HII regions in spiral arms). Indeed after 6 Myr the
H$\alpha$ flux is expected to drop by a factor of $\sim$ 30
while the FUV flux remains almost constant.\\
To check whether a change in the micro history might be at the origin
of the observed variation of the H$\alpha$ over FUV flux ratio, we
simulate the bursty star formation activity of a galaxy by considering
the contribution of single HII regions treated as subsequent
instantaneous bursts of intensity 10$^4$ M$\odot$ yr$^{-1}$ 
repeated every $\Delta T$ years. These toy-models have been computed using the Starburst99 code
(Leitherer et al. 1999; V{\'a}zquez \& Leitherer 2005) with a Kroupa (2001) IMF and
a solar metallicity (but we checked that metallicity has little influence). We then suppose that
the total emission of a galaxy is the sum of the contribution of the
different generations of HII regions\footnote{This assumption is
  reasonable since the diffuse emission of late-type galaxies, that
  can contribute up to 50\% of the total H$\alpha$ emission, is
  probably due to the leakage of ionizing photons from HII regions
  (Oey et al. 2007).}  formed during the time, with the H$\alpha$
emission dominated by the last generation of HII regions, while all
HII regions formed during about the last 100 Myr contribute to the
FUV emission. If, as in the case of massive galaxies, HII regions are
formed at a constant rate with a frequency higher than the life time
of the ionizing OB stars (10$^7$ yrs) (induced star formation),
the resulting H$\alpha$ over FUV ratio is $\sim$ constant and
corresponds to the value determined for a constant star formation
activity. In low luminosity systems, where HII regions are
sporadically formed with a time delay longer than the age of the
ionizing stars,
the observed H$\alpha$ over FUV ratio depends on the age of 
the last burst which only dominates the H$\alpha$ emission of the galaxy.\\

This is well illustrated in Fig. \ref{modsam1}, which shows the
expected variation of the H$\alpha$ over FUV flux ratio as a function
of the age of the last burst for a galaxy with an activity due to
subsequent, instantaneous bursts of star formation happening every
$\Delta T$ years.  

%
Decreasing the frequency of the bursts from $\Delta
T$ = 2 (solid red), 5 (dashed blue) and 10 (dotted green) Myr
increases the H$\alpha$ over FUV flux ratio \emph{at the epoch of the
last burst} just because the contribution of the previous bursts to
the UV emission is less and less important (they were less frequent in
the past for $\Delta T$ = 10 than for $\Delta T$ = 2 Myr). However, the \emph{probability}
of observing galaxies with a lower H$\alpha$ over FUV flux
ratio (theoretical histograms in Fig. \ref{modsam2} are computed assuming every ages of the last
burst within the range 0 to $\Delta T$ have the same probability to occur) 
increases just because of the longer interval of time in between
two bursts. Figure \ref{modsam2} gives the probability distribution
of observing a given $f(H\alpha)/f(FUV)$ flux ratio (in logarithmic
scale) for galaxies with HII regions forming every $\Delta T$ = 2
(solid red), 5 (dashed blue) and 10 (dotted green) Myr.

Figure \ref{modsam2} and Table \ref{Tabstarb} show that the different type of star formation, constant in massive
late-type galaxies while sporadic in low mass systems, can induce on average a decrease of the H$\alpha$ to FUV flux ratio of
$\sim$ 40 \% (from $constant$ to $\Delta T$=10 Myr), a variation more important than that observed in our sample (0.14 dex). 
Furthermore for a sporadic star formation activity ($\Delta T$ $\geq$ 5 Myr) the distribution of the $f(H\alpha)/f(FUV)$ flux ratio
is expected to be more spreaded than for a constant activity, including both high values of H$\alpha$ to FUV flux ratios in the case
of very young episods (ongoing starbursts) and low values whenever the age of the last burst is $\geq$ 2 Myr. While 
in low mass galaxies ($M_{star}$ $\leq$ 10$^{9.2}$ M$\odot$) the observed distribution of Log $f(H\alpha)/f(FUV)$ 
(yellow shaded histogram in Fig. \ref{modsam2}) extends to 
values as low as 0.5 \AA, values higher than the expected ratio for a constant activity (Log $f(H\alpha)/f(FUV)$ $\geq$ 1.2 \AA)
are not present since systematically excluded as starburst galaxies\footnote{Log $f(H\alpha)/f(FUV)$ = 0.87 $\pm$ 0.17 \AA~ for normal 
and 0.99 $\pm$ 0.26 \AA~ for starburst, low mass galaxies.}.  
Values of $\Delta T$ in between 0 and 10 Myr are realistic since they give an average age of HII regions between 0 and 5 Myr, 
consistent with the observations (Helmboldt et al. 2008; Tamburro et al. 2008).

In late-type galaxies the surface brightness and the colour are proxies of morphology. 
The observed relation between $f(H\alpha)/f(FUV)$ and the effective surface brightness $\mu_e(H)$ shown in Fig.
\ref{hafuvmuh} (and colour in Fig. \ref{hafuvfh}) might thus be due to the decreasing relative contribution 
of young to old HII regions to the total H$\alpha$ luminosity of
spiral galaxies of later types, i.e. characterized by a decreasing surface brightness (and bluer colour). 
Being the H band surface brightness and the FUV-H colour correlated with the
galaxy stellar mass (Gavazzi et al. 1996), the same effect might be at the origin of the observed $f(H\alpha)/f(FUV)$
vs. $M_{star}$ relation shown in Fig. \ref{hafuvmstar}.

   \begin{figure}
   \centering
   \includegraphics[width=15cm,angle=0]{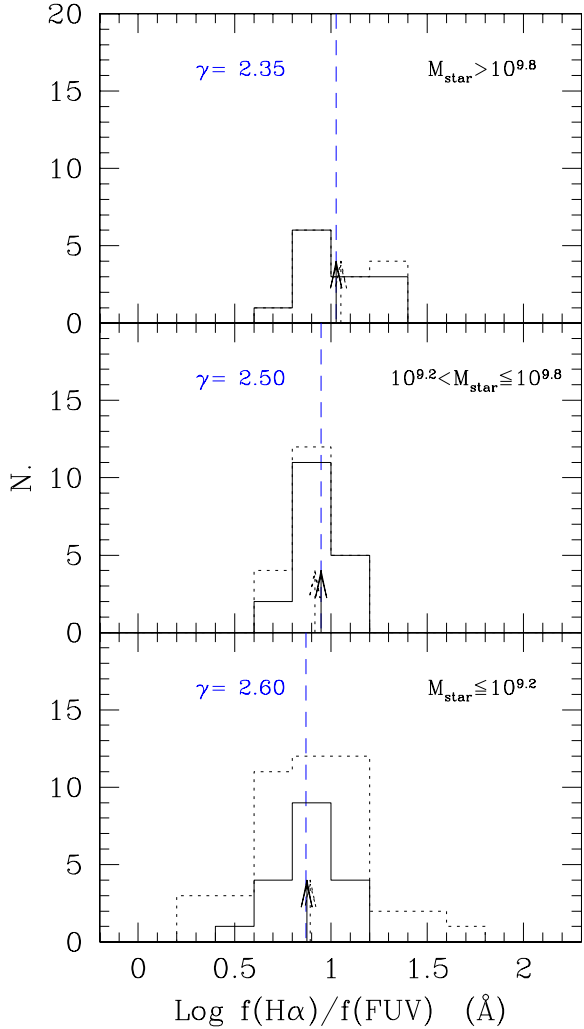}
   \caption{The H$\alpha$ to FUV flux ratio distribution (in logarithmic scale)
   for normal, late-type galaxies with high (solid line) and medium (dotted line)
   quality data in three different intervals in stellar mass:
   massive ($M_{star}$ $>$ 10$^{9.8}$ M$\odot$; upper panel),
   intermediate mass (10$^{9.2}$ $<$ $M_{star}$ $\leq$ 10$^{9.8}$ M$\odot$; central panel)
   and low mass ($M_{star}$ $\leq$ 10$^{9.2}$ M$\odot$; upper panel) galaxies. The vertical arrows indicate
   the mean values of the distributions (see Table \ref{Tabmass}). The vertical dashed line gives the $\gamma$
   slope for a Salpeter IMF.
    }
   \label{distr}
  \end{figure}

\begin{figure}
  \centering
  \includegraphics[width=15cm,angle=0]{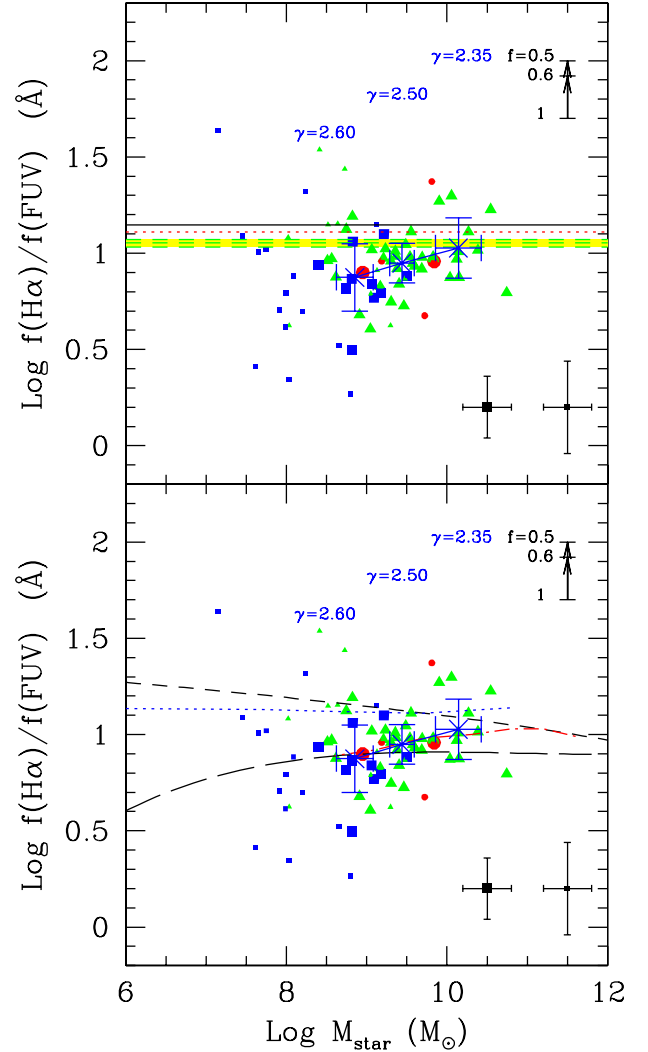}
  \caption{The relationship between the H$\alpha$ to FUV flux ratio
    (in logarithmic scales) and the total stellar mass for ``normal''
    late-type galaxies. Symbols are as in Fig. \ref{hafuvmet}. 
    Data are compared to the prediction of constant (upper panel) and
    variable (lower panel) IMF. The big, blue crosses indicate
    the mean values of Log $f(H\alpha)/f(FUV)$ as determined from our
    high quality sample in three different intervals of stellar mass,
    $M_{star}$ $>$ 10$^{9.8}$, 10$^{9.2}$ $<$ $M_{star}$ $\leq$
    10$^{9.8}$, $M_{star}$ $\leq$ 10$^{9.2}$ M$\odot$. The $\gamma$
    values are the slope of the IMF assuming a mass range between 0.1
    and 100 M$\odot$ (adapted from Meurer et al. (2009)). Upper panel: the
    black solid line gives the expected ratio for the Kennicutt
    (1998) calibrations, the red dotted line for a Kroupa (2001) IMF,
    the yellow shaded region is for a Salpeter ($\gamma$=2.35,
    $M_{up}$=100 M$\odot$) IMF as determined using three different
    stellar population models, where the green, dashed lines are, from
    top to bottom, GALAXEV (Bruzual \& Charlot 2003), Starburst99
    (Leitherer et al. 1999) and PEGASE (Fioc \& Rocca-Volmerange 1997)
    (adapted from Meurer et al. 2009).
    Lower panel:
    the blue dotted line indicates a Bicker \& Fritze-Alvensleben
    (2005) IMF, the black short-dashed line a Leitherer
    (2008) IMF, both determined assuming a metallicity-mass
    relation (see text).  The black, long-dashed line shows the
    expected variation of Log $f(H\alpha)/f(FUV)$ vs. stellar mass
    for the integrated galaxy IMF of Pflamm-Altenburg et al. (2007;
    2009), the blue dotted line for the proposed
    luminosity dependent IMF of Hoversten \& Galzebrook (2009).   }
  \label{hafuvmstarimf} \end{figure}

Finally, we note that a similar
explanation (burst phases separated by long quiescent periods) was
advanced in Boissier et al. (2008) to explain the FUV-NUV colour of
some Low Surface Brightness galaxies (only the delay between bursts
was of the order of 100 Myr to affect FUV-NUV instead of 10 Myr for the H$\alpha$
to FUV flux ratio).

\subsection{Initial mass function}

The determination of the SFR of galaxies using standard recipes as those given by
Kennicutt (1998) resides on the hypothesis that galaxies have a constant initial mass function (IMF).
Several new evidences, however, questioned this general statement. 

The relative number of ionizing to non ionizing stars would decrease in dwarf systems
if these objects have a steeper (or truncated) IMF compared to the Kennicutt one,  
as claimed by Hoversten \& Glazebrook (2008) and Meurer et al. (2009). 
Figures \ref{distr} and \ref{hafuvmstarimf} show that a variation of the slope of the IMF from
$\gamma$ = 2.35 (Salpeter) in massive galaxies down to $\gamma$ = 2.60 (or equivalently a decrease of the upper mass cutoff)
in low mass systems (assuming $f$ = 1) would be enough to reproduce the observed trend. These values are 
consistent with those proposed by Hoversten \& Glazebrook (2008) (red dashed-dotted line in Fig. \ref{hafuvmstarimf})
but very different from those proposed Meurer et al. (2009), where $\gamma$ has to change from $\sim$ 1.5 to
$\sim$ 3.5 (if the upper mass cutoff is kept constant) to reproduce their observed variation in the H$\alpha$ to FUV flux ratio. 
A non universal IMF could thus easily explain the observed trends.\\ 
In massive galaxies ($M_{star}> 10^{9.8}$ M$\odot$), where star formation can be considered as constant, 
Log $f(H\alpha)/f(FUV)$ = 1.03 $\pm$ 0.16 \AA. This value can be compared to the expected ratios for different constant IMF: 
1.07, 1.05 or 1.03 \AA~ for a Salpeter IMF (with $M_{up}$=100 M$\odot$) using respectively
the population synthesis models GALAXEV (Bruzual \& Charlot 2003), Starburst99 (Leitherer et al. 1999) 
and PEGASE (Fioc \& Rocca-Volmerange 1997), 1.15 \AA ~is the value predicted by Kennicutt (1998)
still using a Salpeter IMF and 1.11 \AA ~for Kroupa (2001).
All these predicted values are larger than the observed value in massive galaxies, and are thus consistent with
the observations since $f$ is $\leq$ 1.

   \begin{figure*}
   \centering
   \includegraphics[width=15cm,angle=0]{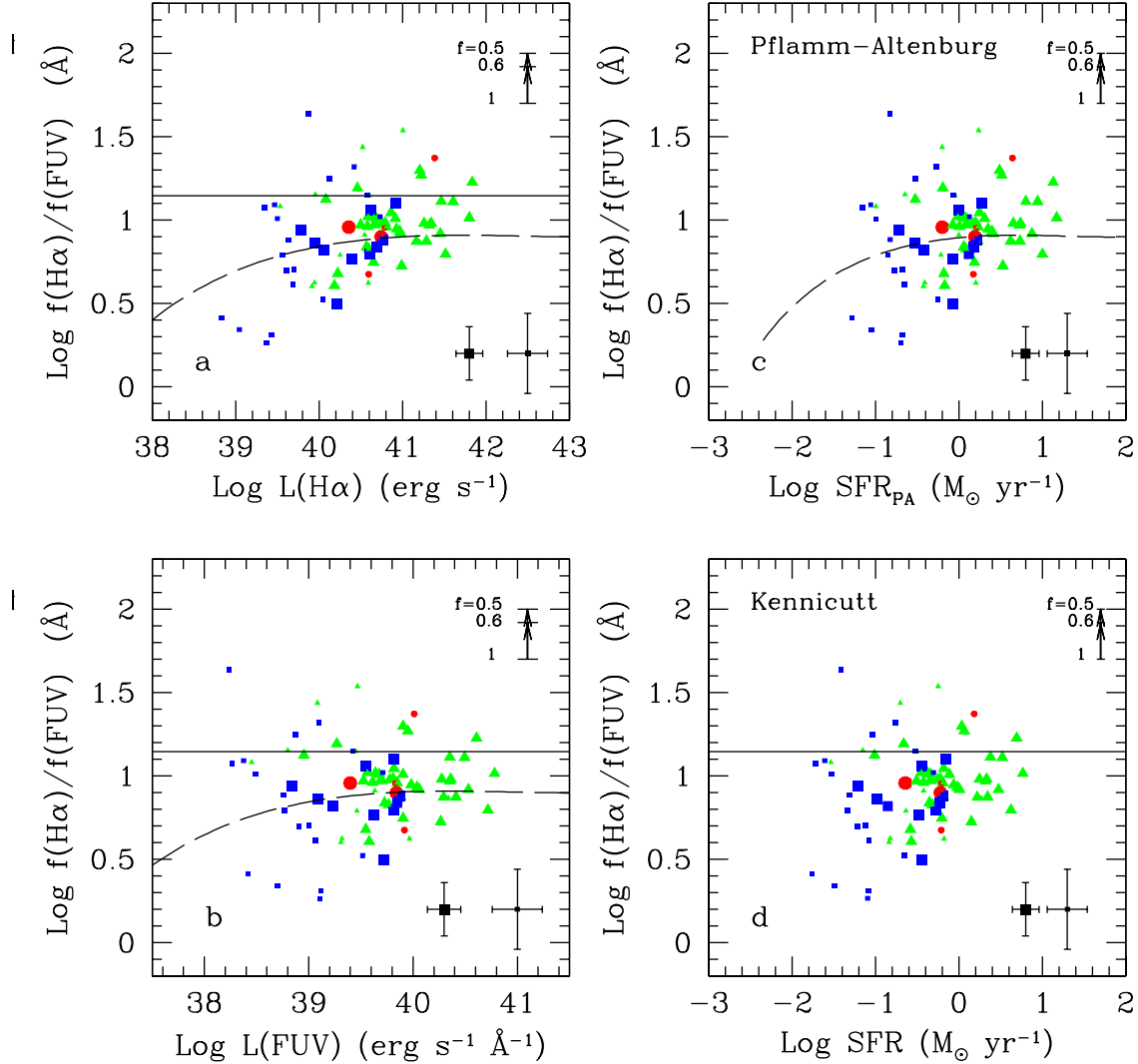}
   \caption{The relationship between the H$\alpha$ to FUV flux ratio 
   and the a) H$\alpha$ luminosity, b) the FUV luminosity, c) the average H$\alpha$ and FUV star formation rate as determined 
   using the luminosity dependent calibration of Pflamm-Altenburg et al. (2007; 2009) and d) 
   the average H$\alpha$ and FUV star formation rate as determined 
   using the Kennicutt (1998) calibration for the normal, late-type galaxies. Symbols are coded as in Fig. \ref{hafuvmet}.
   The black, long dashed line represents the expected variations predicted by Pflamm-Altenburg et al. (2007; 2009) (standard model).
    }
   \label{hafuvsfr}
  \end{figure*}

\subsubsection{Integrated galactic initial mass function}

Pflamm-Altenburg and collaborators, based on statistical considerations, have recently questioned the universality of the IMF when considered as 
an integrated value for a given galaxy (IGIMF), a function appropriate for describing the IMF of unresolved galaxies 
such as those studied in extragalactic astronomy and cosmology. It has been shown
that the maximum stellar mass in star clusters is limited by the mass of the cluster, and that
the mass of the cluster itself is constrained by the star formation rate.
Combining these considerations, Pflamm-Altenburg et al. (2007) have shown that 
the form of the integrated galactic initial mass function IGIMF is not universal but rather
depends on the SFR of the parent galaxy.
They thus concluded that using constant SFR(H$\alpha$) vs. $L(H\alpha)$ calibrations would lead to a systematic and significant (up to 
3 orders of magnitude!) underestimate of the SFR, in particular in dwarf systems.
The non-linear regime of the SFR(H$\alpha$) vs. $L(H\alpha)$ relation is expected
for H$\alpha$ luminosities $L(H\alpha)$ $\leq$ 10$^{40}$ erg s$^{-1}$, the typical range for 
galaxies with $M_{star}$ $\leq$ 10$^{8.5}$ M$\odot$.
Although in a smaller way than for H$\alpha$, the variation of the IGIMF is also expected to affect the 
UV non-ionizing radiation (Pflamm-Altenburg et al. 2009), thus to modify the H$\alpha$ to FUV flux ratio (black, long dashed line
in Figure \ref{hafuvsfr}). The observed weak trends in the data are consistent with the expected variations of 
Log $f(H\alpha)/f(FUV)$ with the H$\alpha$ and FUV luminosities or
with the SFR, here determined using the average between the H$\alpha$ and FUV estimates based on both the
Pflamm-Altenburg et al. (2007; 2009; $SFR_{PA}$) and Kennicutt (1998) calibrations.\\
Using the tight SFR vs. $M_{star}$ relation (SFR (M$\odot$ yr$^{-1}$) = 0.92$\times$ Log$M_{star}$ (M$\odot$) - 8.8,
where the SFR is determined using the standard Kennicutt calibrations) observed in our sample, 
we can see whether the observed $f(H\alpha)$ over $f(FUV)$ flux ratio vs. $M_{star}$ relation shown in Fig. \ref{hafuvmstarimf} is
due to the variation of the IGIMF (long dashed line).
A varying IGIMF should induce a decrease of $f(H\alpha)/f(FUV)$ for galaxies with stellar masses $M_{star}$ $\sim$ $\leq$ 10$^{8.5}$ M$\odot$,
while for galaxies with higher masses $f(H\alpha)/f(FUV)$ should have a constant value of $\sim$ 8 \AA. 
The large dispersion in the data plotted in Fig. \ref{hafuvmstarimf} and \ref{hafuvsfr} and the relatively small
dynamic range in star formation of our target galaxies do not allow us
to discriminate between a constant or a star forming dependent IMF. In massive galaxies the expected value of the H$\alpha$ to FUV flux ratio
for an IGIMF (standard model) is Log $f(H\alpha)/f(FUV)$ = 0.91 \AA, a value lower more than one $\sigma$ from the observed distribution if 
$f$ $<$ 0.91. The IGIMF seems thus to be rejected (1 $\sigma$ confidence level) if more than $\sim$ 10\% of the ionizing radiation is absorbed by dust.

   \begin{figure}
   \centering
   \includegraphics[width=15cm,angle=0]{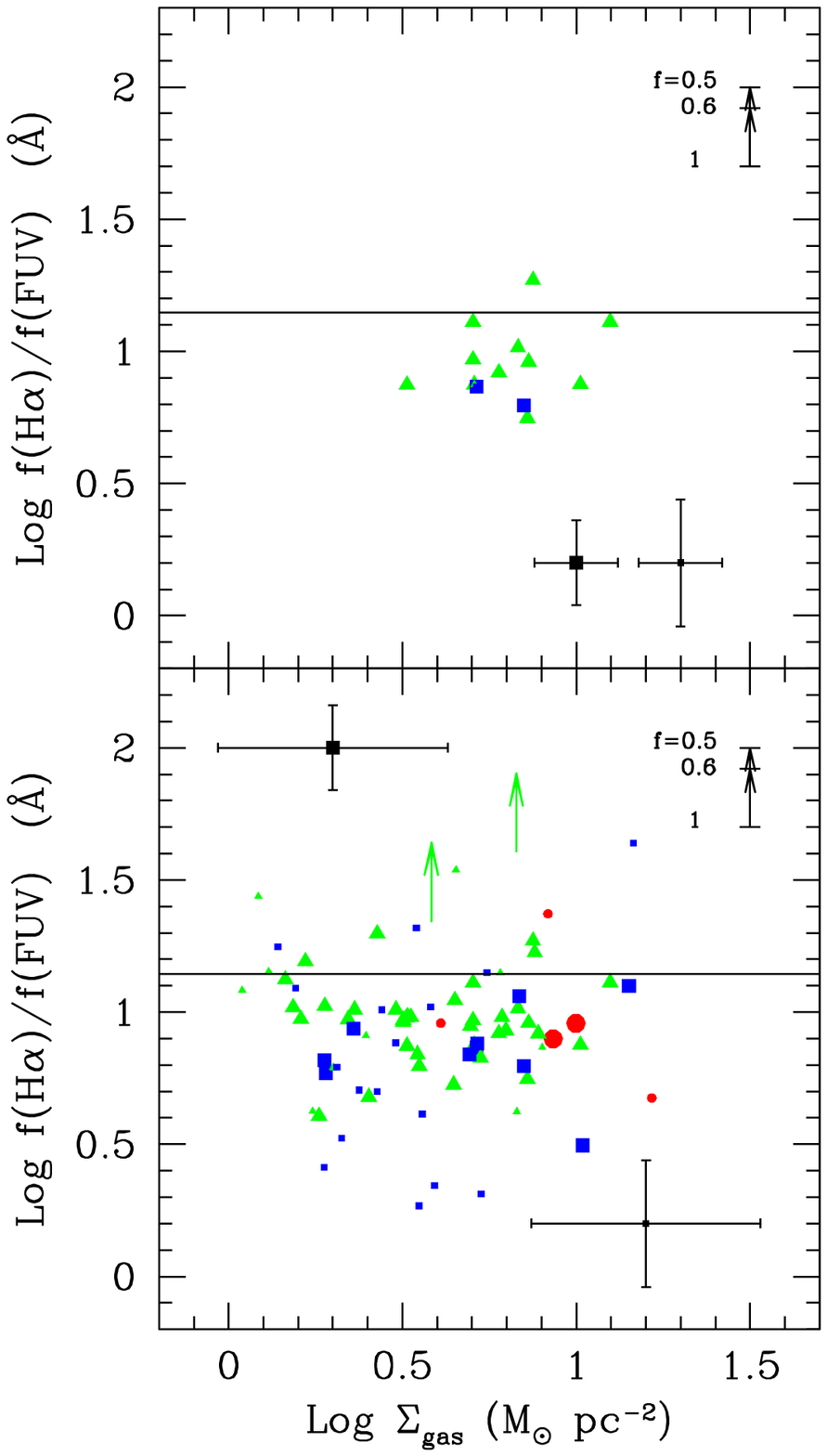}
   \caption{The relationship between the H$\alpha$ to FUV flux ratio 
   and total gas surface density for galaxies with available HI maps (upper panel)
   and for objects where the gas surface density has been inferred using 
   a standard relation between the HI and the optical diameters (lower panel). 
   Symbols are coded as in Fig. \ref{hafuvmet}.
    }
   \label{hafuvsigmagas}
  \end{figure}

\subsubsection{IMF vs. gas column density}

Theoretical considerations lead Krumholz \& McKee (2008) to predict that only gas clouds with
column densities $\geq$ 1 g cm$^{-2}$ can avoid fragmentation and
form massive stars. For this reason they expect a truncation of the IMF in low density regimes
such as the outer discs of spiral galaxies which would almost suppress the
H$\alpha$ emission while only decrease the UV one. Extended UV discs without
H$\alpha$ emission have been indeed observed by GALEX in several nearby 
galaxies (Thilker et al. 2007, Boissier et al. 2007). Krumholz \& McKee (2008)
claim that for this reason a possible decrease of the $f(H\alpha)/f(FUV)$ ratio could be also expected in those objects
not dominated by giant clouds such as dwarf galaxies. \\
To test this hypothesis we 
see whether the $f(H\alpha)/f(FUV)$ ratio is related to the total gas column density of our sample galaxies
with the assumption that the total gas column density scales as the local gas density,
which in star forming regions is significantly higher.  
Gas profiles are available only for a bunch
of objects from the HI surveys of Warmels (1986) and Cayatte et al. (1994).
By comparing their optical and HI distributions, however, we determined a 
general relation linking their optical to their HI diameter.
Figure \ref{hafuvsigmagas} shows the relationship between the $f(H\alpha)/f(FUV)$ ratio
and the observed (upper panel) and inferred (lower panel) total gas surface densities of the target galaxies.

\noindent
The data do not show any relationship between the H$\alpha$ to FUV ratio 
and the total gas column density. Because of gas dilution on the scale of galactic discs, however, the measured
gas column densities are a factor of $\sim$ 10$^3$ smaller than those needed for high mass star 
formation ($\geq$ 1 g cm$^{-2}$). It is thus likely that the total gas column density here determined is not a good proxy of
the local gas density. High resolution data are thus needed to check the validity of this theory.

\subsubsection{IMF vs. redshift}

Galaxy formation models based on cold gas accretion (Brinboim \& Dekel 2003; Keres et al. 2005; Dekel et al. 2009)
predict that the star formation rate is a fairly steady function of time (e.g. Finlator et al. 2006; Finlator et al. 2007), 
with a tight relationship between stellar mass and star formation activity, slightly evolving with redshift.
Observations (e.g. Noeske et al. 2007; Elbaz et al. 2007; Daddi et al. 2007), however, revealed that the amplitude of 
the $M_{star}$ vs. SFR relation evolve with time in a way inconsistent with model predictions. Going to 
high redshift the observed SFRs are higher and/or the stellar masses are lower than predicted by models (Dav\'e 2008; Wilkins et al. 2008).
To justify this observational evidence several authors proposed that the IMF is not universal but rather evolving with time, being
top heavy at high redshift (Larson 1998; Baugh et al. 2005; Dav\'e 2008; Lacey et al. 2008; Wilkins et al. 2008) when the universe was 
dominated by starbursts (Rieke et al. 1993; Parra et al. 2007). Given the large uncertainties in the dust extinction correction for starbursts,
the present analysis can not be used to infer any variation of the IMF with cosmic time.

\section{Conclusion}

We have studied the high mass star formation activity of an optically/near-IR selected sample of late-type galaxies using two 
independent, direct tracers of star formation, the H$\alpha$ and the FUV luminosity. This analysis
brought the following results:\\

1) The distribution of the H$\alpha$ over FUV flux ratio  
is highly dispersed (Log $f(H\alpha)/f(FUV)$ = 1.10 $\pm$ 0.34 \AA) when the data are properly corrected for dust attenuation using 
the Balmer decrement and the TIR to FUV flux ratio. The dispersion increases (Log $f(H\alpha)/f(FUV)$ = 1.00 $\pm$ 0.41 \AA) when
are included data corrected using statistical recipes. This result indicates that H$\alpha$ and UV
luminosities, if blindly combined with standard recipes as those given by Kennicutt (1998), give star formation rates with at best an uncertainty of
a factor of $\sim$ 2-3 depending on the applied dust attenuation correction. 
The error is significantly reduced (Log $f(H\alpha)/f(FUV)$ = 0.94 $\pm$ 0.16 \AA) once AGN, 
starburst and highly inclined late-type galaxies are excluded.\\

2) Highly inclined galaxies, generally characterized by prominent dust lanes, have H$\alpha$ over FUV flux ratio  
significantly higher (Log $f(H\alpha)/f(FUV)$ = 1.44 $\pm$ 0.32 \AA) than normal, late-type galaxies. This result suggests that the 
dust attenuation corrections applied to this particular class of objects is wrong, probably because the assumption of isotropy in the UV emission of the
galaxy is not correct (Panuzzo et al. 2003) or because the Balmer decrement does not give an accurate estimate of the extinction (photon leakage and/or porosity
in the ISM). \\

3) Extinction is the major source of uncertainty in the determination of the H$\alpha$ to FUV flux ratio even in normal, late-type galaxies.
A direct measure of the Balmer decrement and, to a lesser extent, of the TIR/FUV flux ratio are mandatory for avoiding any residual,
systematic effect in the determination of the emitted H$\alpha$ and FUV fluxes, thus on the star formation activity of the target galaxies.\\

4) The data are consistent with a Kennicutt (1998), Kroupa (2001) and Salpeter IMF. To match the data, 
the fraction of ionizing photons absorbed by dust (1-$f$) should be smaller than previously estimated ($f$=0.57, Hirashita et al. 2003), with $f$ ranging
from $\sim$ 1 for a Salpeter IMF to 0.77 for Kennicutt (1998). The star formation dependent IGIMF proposed by Pflamm-Altenburg et al. (2007; 2009)
can be hardly constrained because of the low dynamic range in star formation
of our sample. We can only state that the proposed IGIMF is rejected (at 1 $\sigma$ level) in massive galaxies if $f$ $<$ 0.91.\\

5) The H$\alpha$ over FUV flux is barely related with stellar mass, FUV-H colour, H band effective surface brightness, 
star formation rate and gas fraction in normal, late-type galaxies. The steepness of these relationships increases if a mass/metallicity dependent $f$
factor is assumed. With the present dataset it is impossible to state whether these trends are real or due to a residual in the adopted
extinction corrections. If real, these trends can be explained by a small variation in the IMF, with a slope 
$\gamma$ $\sim$ 2.35 in massive systems and 
$\gamma$ $\sim$ 2.60 in dwarfs if $f$=1, while larger variations should be invoked if $f$$<$ 1 in massive galaxies. We have however 
shown that the different micro history of star formation in giant,
high surface brightness galaxies, characterized by induced star formation along their spiral arms, with respect to dwarf, 
low surface brightness systems where the ionizing and non ionizing UV emission is dominated by sporadically formed, 
giant HII regions can explain all the observed trends previously described. \\

The consequences of these results are major not only in the study of
the star formation properties of nearby galaxies but also in a
cosmological context as, for example, in the determination of the cosmic
variation of the star formation activity of the universe. We indeed
give an accurate estimate of the error on the star formation activity
of galaxies using standard recipes based on two among the most widely
used tracers, Balmer lines and UV luminosities. Variations in the IMF,
as those claimed by Meurer et al. (2009) and Hoversten \& Galzebrook
(2008), which would have drastic consequences in the determination of
the mass assembly through stellar formation during the evolution of
the universe, are no more required to explain the observations.
This micro history variation
scenario, here proposed to explain the observed trends, should be tested in the study of the HII distribution of
different nearby galaxies spanning a large range in morphological
type, luminosity and surface brightness. If confirmed, its more
direct consequence would be that the star formation activity of dwarf
galaxies, although almost constant on long time scales (Sandage 1986;
Gavazzi et al 1996; Boselli et al. 2001), is bursty on short times
scales. The stationarity condition required for
transforming H$\alpha$ and UV luminosities into star formation rates
would thus not be satisfied in these low luminosity systems, for which
other techniques such as SED fitting would be required for measuring
their activity.

\acknowledgements 
We want to thank JM Deharveng for interesting discussions and the anonymous referee for precious comments
which helped increasing the quality of the manuscript. 
This research has made use of the NASA/IPAC Extragalactic Database (NED) 
which is operated by the Jet Propulsion Laboratory, California Institute of 
Technology, under contract with the National Aeronautics and Space Administration. 
We acknowledge the usage of the HyperLeda database (http://leda.univ-lyon1.fr)
and the GOLDMine database (http://goldmine.mib.infn.it/).

\newpage

\begin{table*}
\caption{The data completeness in the sample}
\label{Tabcomp}
{\scriptsize
\[
\begin{tabular}{cccc}
\hline
\noalign{\smallskip}
Sample 		&   Balmer~dec+TIR/UV	 & Balmer~dec	 & TIR/UV	\\
\hline
HRS$^a$		&   40  		 &	 46	 &  63	\\
Virgo		&   36  		 &	 65	 &  96	\\
Atlas		&   35  		 &	 37	 &  39	\\
\hline
Total		&   111 		 &	 148	 & 198	\\
\noalign{\smallskip}
\hline
\end{tabular}
\]
}
Notes: \\
a: excluding HRS galaxies in the Virgo cluster\\
\end{table*}

\begin{table*}
\caption{The three quality samples}
\label{Tabsamp}
{\scriptsize
\[
\begin{tabular}{ccccc}
\hline
\noalign{\smallskip}
Sample 		& N.~galaxies	& error~on~Log~f(H$\alpha$)/f(FUV)& Selection~criterion		& Extinction~correction\\
\hline
High~quality	& 111(49)		& 0.16				& Balmer~decrement~+~TIR/FUV  	& H$\alpha$/H$\beta$~+~TIR/UV\\
Medium~quality	& 148(81)		& $\leq$ 0.24			& Balmer~decrement		& H$\alpha$/H$\beta$~+~colour~dependent~statistical~corrections\\
Low~quality	& 198(120)		& $\leq$ 0.34			& 				& A(H$\alpha$) vs.M$_{star}$ ~+~colour~dependent~statistical~corrections\\
\noalign{\smallskip}
\hline
\end{tabular}
\]
}
Notes: the medium and low quality samples include also the higher quality samples. Values in parenthesis are for 
``normal'' galaxies.
\end{table*}

\begin{table*}
\caption{The mean values of the H$\alpha$ to FUV flux ratio}
\label{Tabratio}
{\scriptsize
\[
\begin{tabular}{ccccccc}
\hline
\noalign{\smallskip}
Sample	 	& H.Q.  &		   		&	M.Q.	&		   & L.Q.&		  \\
\hline
	 	& N.	&$<$Log $f(H\alpha)/f(FUV)$ $>$ (\AA)&	N.	&$<$Log $f(H\alpha)/f(FUV)$ $>$ (\AA)& N.&$<$Log $f(H\alpha)/f(FUV)$ $>$ (\AA)\\
\hline
All	 	& 111	& 1.10 $\pm$ 0.34  		& 148		& 1.05 $\pm$ 0.37  		& 198	& 1.00 $\pm$ 0.41 \\
AGN	 	& 23	& 1.23 $\pm$ 0.34  		& 24		& 1.28 $\pm$ 0.42  		& 32	& 1.34 $\pm$ 0.40 \\
Starburst	& 33	& 1.15 $\pm$ 0.44  		& 33		& 1.15 $\pm$ 0.44  		& 35	& 1.16 $\pm$ 0.44 \\
incl$>$80$^O$ $^a$&  9	& 1.44 $\pm$ 0.32  		& 13		& 1.23 $\pm$ 0.44  		& 15	& 1.11 $\pm$ 0.52 \\
Normal   	& 49	& 0.94 $\pm$ 0.16  		& 81		& 0.93 $\pm$ 0.25  		& 120	& 0.87 $\pm$ 0.32 \\
\noalign{\smallskip}
\hline
\end{tabular}
\]
}
Notes: a) including all highly inclined galaxies listed in Table \ref{Tabestremi}.
\end{table*}

\begin{table*}
\caption{Galaxies with a Log $f(H\alpha)/f(FUV)$ $>$ 1.35 \AA$^a$}
\label{Tabestremi}
{\scriptsize
\[
\begin{tabular}{cccccc}
\hline
		&		&		&Highly~inclined~galaxies	&	&		\\
\noalign{\smallskip}
\hline
Name		& type 		& Log $M_{star}$& Log $f(H\alpha)/f(FUV)$	& incl	& Comments     \\
\hline
		&		& M$\odot$	& \AA				& $^o$	&		\\
\hline
NGC3424		& SB(s)b:?;HII	& 9.81		& 1.70				& 78	& Interacting system \\
NGC3683		& SB(s)c?;HII	&10.14		& 1.70				& 69	& With prominent dust lane \\
NGC4517		& SA(s)cd: sp	&10.32		& 1.50				& 81	& With prominent dust lane \\
NGC4631		& SB(s)d 	& 9.92		& 1.68				& 82	& With prominent dust lane \\
NGC5348		& SBbc: sp	& 9.09		& 1.55				& 90	& With prominent dust lane \\
IC1048		& S		& 9.72		& 1.72				& 74	& 			   \\
\noalign{\smallskip}
\hline
		&		& 		& Starburst~galaxies		&		&		\\
\noalign{\smallskip}
\hline
Name		& type 		& Log $M_{star}$& Log $f(H\alpha)/f(FUV)$	& F60/F100	& Comments     \\
\hline
		&		& M$\odot$	& \AA				& 		&		\\
\hline
NGC2798		& SB(s)a pec;HII& 9.87		& 3.08				& 0.70		& Interacting system \\ 
M82		& I0;Sbrst;HII	& 9.93		& 1.61				& 0.94		& Interacting system \\
NGC4383		& Sa? pec;HII	& 9.52		& 1.48				& 0.66		& With peculiar H$\alpha$ tails \\
UGCA284		& BCD:		& 8.41		& 1.41				& 0.53		& Single HII region  \\
NGC5145		& S?;HII;Sbrst	& 9.47		& 1.59				& 0.51		&                    \\
NGC7771	&SB(s)a;HII;LIRGSbrst	&11.02		& 1.42				& 0.55		& Interacting system \\
\hline
\end{tabular}
\]
}
Notes: a = Log $f(H\alpha)/f(FUV)$ $=$ 1.35 \AA ~ is 2 $\sigma$ higher than the mean value of normal, massive galaxies.
\end{table*}
  
\begin{table}
\caption{The H$\alpha$ to FUV flux ratio for normal, late-type galaxies of different mass}
\label{Tabmass}
{\scriptsize
\[
\begin{tabular}{cccc}
\hline
\noalign{\smallskip}
Sample 							&quality& N.~obj.	& $<$ Log $f(H\alpha)/f(FUV)$ $>$ \AA	\\
\hline
$M_{star}$ $>$ 10$^{9.8}$ M$\odot$			&high	& 13 		& 1.03	$\pm$ 0.16	\\
							&medium	& 14		& 1.05	$\pm$ 0.18	\\
10$^{9.2}$ $<$ $M_{star}$ $\leq$ 10$^{9.8}$ M$\odot$	&high	& 18 		& 0.95	$\pm$ 0.10	\\
							&medium	& 21		& 0.93	$\pm$ 0.13	\\
$M_{star}$ $\leq$ 10$^{9.2}$ M$\odot$			&high	& 18 		& 0.87	$\pm$ 0.17	\\
							&medium	& 46		& 0.89	$\pm$ 0.30	\\
\noalign{\smallskip}
\hline
\end{tabular}
\]
}
\end{table}

 \begin{table}
\caption{The H$\alpha$ to FUV flux ratio vs. Log $M_{star}$ relation using data corrected according to different recipes}
\label{Tabcor}
{\scriptsize
\[
\begin{tabular}{ccccc
}
\hline
\noalign{\smallskip}
Extinction      	& correction	&		 &			&	\\
\hline
\noalign{\smallskip}
H$\alpha$ 		& FUV		&	a	 & b			& R$^c$\\
\hline
Observed		& Observed	& 0.13$\pm$0.05	 & -0.03$\pm$0.42	& 0.40 \\
H$\alpha$/H$\beta$(TW)	& FIR/UV(TW)	& 0.08$\pm$0.04  &  0.18$\pm$0.37  	& 0.29 \\
H$\alpha$/H$\beta$(TW)	& Calzetti	&-0.03$\pm$0.03  &  1.17$\pm$0.30  	&-0.08 \\
Statistical$^d$		& FIR/UV(TW)	& 0.35$\pm$0.03  & -2.27$\pm$0.30  	& 0.85 \\
H$\alpha$/H$\beta$(TW)	& Statistical$^d$& 0.11$\pm$0.05  & -0.15$\pm$0.43  	& 0.32 \\
Statistical$^d$		& Statistical$^d$& 0.38$\pm$0.04  & -2.60$\pm$0.30  	& 0.80 \\
\hline
\end{tabular}
\]
}
Notes: linear fit to the data: Log $f(H\alpha)/f(FUV)$ = a(X-variable) + b
for the subset of 49 normal galaxies of the high quality sample.\\
c: linear correlation coefficient.\\
d: statistical corrections as those applied by Meurer et al. (2009).\\
\end{table}

\begin{table}
\caption{The $f$ values for different IMF}
\label{Tabf1}
{\scriptsize
\[
\begin{tabular}{ccc}
\hline
\noalign{\smallskip}
IMF 						&Log $f(H\alpha)/f(FUV)$ \AA& $f$	\\
\hline
Kroupa 2001					& 1.11     	    & 0.84  \\
Salpeter (0.1$<m<$100 M$\odot$) (Kennicutt 1998)& 1.15     	    & 0.77  \\
Salpeter (0.1$<m<$100 M$\odot$) (Starburst99)	& 1.05     	    & 0.95  \\
Salpeter (0.1$<m<$100 M$\odot$) (PEGASE)	& 1.03     	    & 0.99  \\
Salpeter (0.1$<m<$100 M$\odot$) (GALAXEV)	& 1.07     	    & 0.91  \\
\noalign{\smallskip}
\hline
\end{tabular}
\]
}
\end{table} 
 
\begin{table}
\caption{The best fit to the data}
\label{Tabfit}
{\scriptsize
\[
\begin{tabular}{cccccc}
\hline
High quality sample		&		&		&	&		&   \\
\hline
Log $f(H\alpha)/f(FUV)$	\AA	&		 &		  &	  &	      &	     \\
\hline
\noalign{\smallskip}
X-variable	   		& a     	 & b     	  & R$^c$ &$\rho$ $^d$& \%$^e$\\
\hline
Log $M_{star}$	   		& 0.08$\pm$0.04  & 0.18$\pm$0.37  & 0.29  & 0.27	& 93.7 \\
FUV-H		   		& 0.09$\pm$0.03  & 0.73$\pm$0.07  & 0.40  & 0.39   	& 99.3 \\
$\mu_e(H)$	   		&-0.08$\pm$0.03  & 2.50$\pm$0.54  & 0.39  &-0.37	& 98.9 \\
12+Log(O/H)	   		& 0.05$\pm$0.14  & 0.51$\pm$1.22  & 0.05  & 0.02 	& 10.5 \\
Log $M_{gas}/(M_{gas}+M_{star}$)&-0.36$\pm$0.09  & 0.81$\pm$0.04  & 0.50  &-0.40  	& 99.5 \\
Log $SFR$			& 0.08$\pm$0.05  & 0.96$\pm$0.02  & 0.21  & 0.21	& 85.7 \\
\noalign{\smallskip}
\hline
\hline
Medium quality sample		&		&		&	&		&   \\
\hline
Log $f(H\alpha)/f(FUV)$	\AA	&		 &		  &	  &	      &	     \\
\hline
\noalign{\smallskip}
X-variable	   		& a     	 & b     	  & R$^c$ &$\rho$ $^d$& \%$^e$\\
\hline
Log $M_{star}$	   		& 0.06$\pm$0.04  & 0.37$\pm$0.37  & 0.19  & 0.21	& 90.8 \\
FUV-H		   		& 0.09$\pm$0.04  & 0.73$\pm$0.09  & 0.29  & 0.35   	& 99.6 \\
$\mu_e(H)$	   		&-0.09$\pm$0.03  & 2.73$\pm$0.60  & 0.35  &-0.39	& 99.8 \\
12+Log(O/H)	   		& 0.13$\pm$0.11  &-0.21$\pm$0.98  & 0.14  & 0.10 	& 59.0 \\
Log $M_{gas}/(M_{gas}+M_{star}$)&-0.39$\pm$0.12  & 0.81$\pm$0.04  & 0.38  &-0.39  	& 99.8 \\
Log $SFR$			& 0.10$\pm$0.05  & 0.97$\pm$0.03  & 0.24  & 0.21	& 90.8 \\
\noalign{\smallskip}
\hline

\end{tabular}
\]
}
Notes: linear fit to the data: Log $f(H\alpha)/f(FUV)$ = a(X-variable) + b
\\
c: linear correlation coefficient.\\
d: generalized Spearman's  rank  order correlation  coefficient.\\
e: Spearman's probability that the two variables are correlated.\\
\end{table}

\begin{table}
\caption{The expected $f(H\alpha)/f(FUV)$ distribution for subsequent starbursts}
\label{Tabstarb}
{\scriptsize
\[
\begin{tabular}{cccc}
\hline
\noalign{\smallskip}
$\Delta T$	& $<$age$>$	& $<f(H\alpha)/f(FUV)>$	& $<$ Log $f(H\alpha)/f(FUV)$ $>$     \\
\hline
Myr		& Myr		& \AA			& \AA				\\
\hline
constant	& -		& 12.8			& 1.11				\\
2		& 1.05$\pm$0.55 & 12.5$\pm$1.7		& 1.09$\pm$0.06			\\
5		& 2.55$\pm$1.41 & 11.8$\pm$5.6		& 1.01$\pm$0.25			\\
10		& 5.05$\pm$2.86 &  9.0$\pm$10.1		& 0.49$\pm$0.74			\\
\noalign{\smallskip}
\hline
\end{tabular}
\]
}
Notes: for a Kroupa (2001) IMF with solar metallicity.
\end{table}

\clearpage

\appendix

\section{A: The observational data}

The dataset used in the present analysis is composed of imaging and spectroscopic data covering the
whole UV to IR spectral range.
H$\alpha$+[NII] narrow band imaging of the HRS late-type galaxies has been recently obtained at the 2.1m 
San Pedro Martir telescope. Data for galaxies within the Virgo cluster region are available from
our previous observations (Boselli \& Gavazzi 2002; Boselli et al. 2002a; Gavazzi et al. 1998, 2002,
2006a), while those for the GALEX UV atlas of galaxies from different sources.\\
UV data have been taken thanks to GALEX in the two UV bands FUV ($\rm
\lambda_{eff}=1539\AA, \Delta \lambda=442\AA$) and NUV ($\rm
\lambda_{eff}=2316\AA, \Delta \lambda=1069\AA$). Observations have been 
taken as part of the Nearby Galaxy Survey (NGS; Gil de Paz et al. 2007), 
the Virgo cluster (Boselli et al. 2005), 
the All Imaging Survey (AIS) or as pointed observations in
open time proposals. \\
Photometric data for the HRS and the GALEX atlas of galaxies have been taken from different sources
in the literature. For all galaxies near-IR data are available from 2MASS (Jarrett et al. 2003).
For the Virgo cluster region
H (1.65 $\mu$m) and K (2.1 $\mu$m) band frames have 
been obtained during a near-IR imaging survey 
(Boselli et al. 1997; 2000; Gavazzi et al. 2000a, 2001).
B and V frames are available for most of the analyzed galaxies 
thanks to our own observations (Gavazzi \& Boselli 1996; 
Gavazzi et al. 2001; 2005a and Boselli et al. 2003).
Fluxes at all wavelengths, including UV and H$\alpha$, were obtained using our own procedures by integrating all
images within elliptical annuli of increasing diameter up to the
optical B band 25 mag arcsec$^{-2}$ isophotal radii. \\
A low resolution ($R$ $\sim$ 1000), integrated spectroscopic survey in the wavelength range
3500-7200 \AA~ of the HRS is under way at the 1.93m OHP telescope.
In order to sample the spectral properties of the whole galaxy, and not just those of
the nuclear regions, observations have been
executed using the drifting technique described in Kennicutt (1992). 
Exposures are taken while constantly and uniformly drifting the spectrograph slit
over the full extent of the galaxy. 
A resolution $R$ $\sim$ 1000
is mandatory for resolving [NII] from H$\alpha$
and estimating the underling Balmer absorption under H$\beta$.
These data are combined with those available for the Virgo cluster region (Gavazzi et al. 2004) or  
for the UV GALEX atlas (Kennicutt 1992;
Jensen et al. 2001; Mosutakas \& Kennicutt 2006).\\
HI observations, needed for excluding the perturbed cluster galaxy population and for studying the
relationship between the high mass star formation activity and the total gas content,
are available for almost all the late-type galaxies 
in the HRS (Springob et al. 2005), in the Virgo cluster region (Gavazzi et al. 2005b) and in
the UV GALEX atlas (Leda database).
A $^{12}$CO(1-0) survey of the HRS galaxies, necessary for determining 
their molecular gas content, is under way at the 12m Kitt Peak telescope. These data are combined
with those available in the literature from the FCRAO survey of Young et al. (1995), 
Kenney \& Young (1988) and Boselli et al. (1995; 2002b) for the Virgo cluster region.\\
Far-infrared IRAS data at 60 and 100 $\mu$m, necessary for correcting UV and optical data for
internal extinction, have been taken
from different sources in the literature.\\
Most of data used in the present analysis relative to the Virgo cluster are available on the 
net on the GOLDMine database (Gavazzi et al. 2003).

\section{B: The derived parameters}

This dataset is used to derive different useful parameters of the selected galaxies: the ionizing and non-ionizing UV radiation,
the stellar mass, the metallicity and the star formation rate (SFR). These entities can be determined only after applying some
corrections to the observed data.\\

   \begin{figure}
   \centering
   \includegraphics[width=15cm,angle=0]{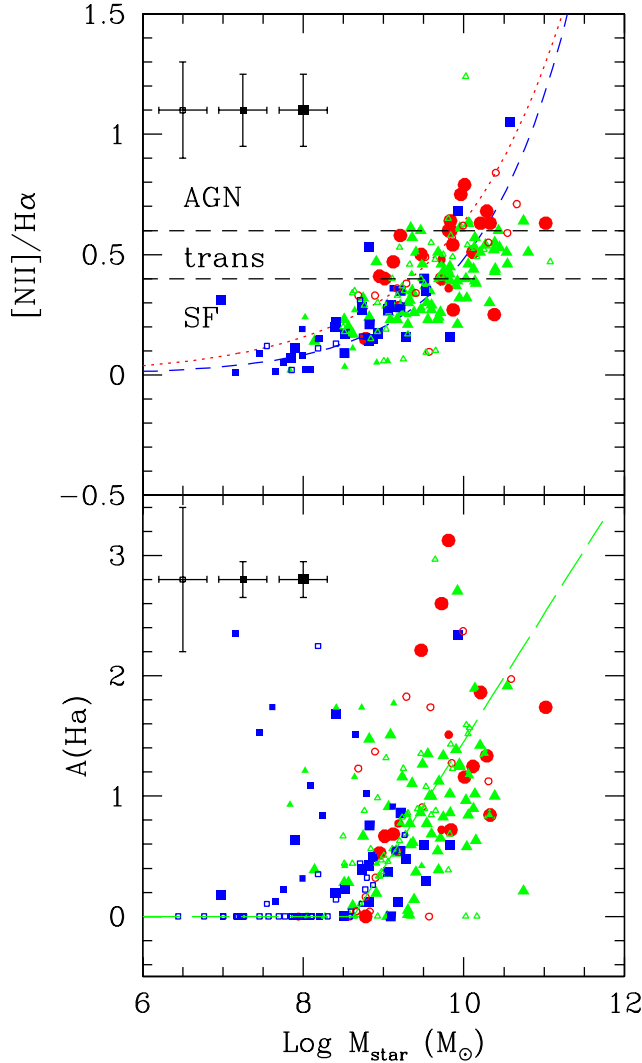}
   \caption{The relationship between a) the [NII]/H$\alpha$ ratio and b) $A(H\alpha)$ and the
   stellar mass (in logarithmic scale). The horizontal, black dashed lines indicate the intervals in [NII]/H$\alpha$
   for AGN, transition and star forming galaxies. The red dotted line gives the empirical 
   [NII]/H$\alpha$  to stellar mass calibration given in Boselli et al. (2002b), the blue short-dashed line
   that of Decarli et al. (2007), while the green long-dashed line the $A(H\alpha)$ vs. Log $M_{star}$
   relation. The Decarli and the $A(H\alpha)$-$L_H$ relations are 
   used in this work whenever spectroscopic data are not available (low quality sample). 
   Symbols are coded as in Fig. \ref{hafuvmet}.
    }
   \label{niiha}
  \end{figure}

{\bf 1) Extinction corrections: }
both H$\alpha$ and UV data must be corrected for dust extinction.
The attenuation in the H$\alpha$ line ($A(H\alpha)$) can be measured using the Balmer decrement determined
from the spectroscopic data. The high resolution ($R$ $\sim$ 1000) and high sensitivity
of the data allow us to make an accurate determination of the underlying Balmer absorption under H$\beta$,
a critical entity necessary for a correct estimate of the Balmer decrement 
(Kobulnicky et al. 1999; Gavazzi et al. 2004). The blending of the [NII] doublet prevents us
from measuring the underlying absorption under H$\alpha$. We thus use a standard correction of 1.5 \AA ~(Gavazzi et al. 2004).
Upper limits are determined for those galaxies with H$\alpha$ line emission and undetected H$\beta$.
The integrated nature of the spectroscopic data is optimized for getting values representative of the 
whole galaxy, avoiding any systematic effect do to aperture corrections necessary when spectroscopic nuclear data are used. 
The attenuation of the H$\alpha$ line is then measured using the galactic extinction law of Lequeux et al. (1979),
as in Cortese et al. (2006). For galaxies 
without spectroscopic data we use a luminosity dependent extinction relation
as determined for those galaxies with available data, plotted in Fig. \ref{niiha} once the H band luminosity is transformed
into stellar masses computed as below:

$A(H\alpha)$ (mag) =1.22$\times$Log $L_H$ (solar units) - 11.28

Given the huge dispersion in this relation, we consider as highly inaccurate those H$\alpha$ 
fluxes corrected without a direct measure of the Balmer decrement (low quality sample).
The H$\alpha$ flux should also be corrected to take into account that a fraction $(1-f)$ of the Lyman continuum photons 
is absorbed by dust before ionizing the atomic hydrogen. For consistency with other works we do not 
apply any $f$ correction.\\
The UV attenuation ($A(UV)$) is determined using the recipes of Cortese et al. (2008) which are based
on the idea that the UV radiation absorbed by dust is re-emitted in the far-IR. For those galaxies without far-IR 
data, $A(UV)$ is measured using the colour or surface brightness dependent relations given in Cortese et al. (2008).
While the use of an energetic balance between the UV absorbed light
and the far-IR emitted radiation is at present the most accurate method for determining the 
UV dust attenuation in galaxies, the use of a colour or surface brightness dependent relation is quite indirect
and still highly uncertain.

{\bf 2) [NII] contamination in H$\alpha$+[NII] imaging data: }
narrow band  H$\alpha$+[NII] imaging data,
used for determining the total H$\alpha$ flux of the selected galaxies,
are corrected for [NII] contamination using the H$\alpha$/[NII] ratio measured from spectroscopic data.
For those galaxies without any spectroscopic information, the H$\alpha$/[NII] ratio is
determined assuming the luminosity dependent relation given in Decarli et al. (2007) and consistent with Boselli et al. (2002b). 
H$\alpha$ fluxes statistically corrected 
for [NII] contamination are more uncertain than those 
based on spectroscopic data. The use of narrow band images and integrated spectra minimize 
aperture effects.

{\bf 3) Stellar mass:}
Stellar masses are calculated using near-IR-optical colour dependent luminosity-mass relations as
determined from our chemo-spectro-photometric models of galaxy evolution (Boissier \& Prantzos 2000).
The availability of near-IR magnitudes for almost all galaxies in the sample,
a direct tracer of both the bulk of the stellar population and of the total dynamical mass
of galaxies (Gavazzi et al. 1996), secure an accurate determination of the total stellar mass
of the selected galaxies.
The adopted relation, computed assuming a Kroupa (2001) IMF, is (with Johnson magnitudes in Vega system):\\

Log $M_{star}$ (M$\odot$) - Log $L_H$ (L$_H$$\odot$) = -1.08 +0.21(B-H)\\

We note that the grid of models was recomputed using the Kroupa (2001) IMF
(rather than the Kroupa et al. (1993) IMF as in Boissier \& Prantzos 1999).
This changes seems to improve the agreement with recent UV data in
SINGS galaxies without producing huge modifications to other results
(Mu\~noz-Mateos et al., in preparation).

{\bf 4) Metallicity:}
Extinction corrected optical emission lines available from integrated spectra have been used to determine the 
metallicity of the observed galaxies using four different calibrations: van Zee et al. (1998), 
Dopita et al. (2000), Kobulnicky et al. (1999) and McGaugh (1991).
The value used here is the mean value determined using these different calibrations.

{\bf 5) Star Formation Rates (SFR):}
H$\alpha$ and UV luminosities can be converted into star formation rates (in M$\odot$ yr$^{-1}$) 
using population synthesis models combined with some hypothesis concerning the star formation history,
the IMF and the metallicity of the observed galaxies. For a given star formation history,
metallicity and IMF, the relationship between the high mass star formation rate and the H$\alpha$ 
and UV luminosities depends on the adopted population synthesis model.
Here SFR are measured using the standard recipes of Kennicutt (1998) unless quoted differently,
 and are average values as determined using H$\alpha$ and FUV data.

\section{C: Error budget}

Observational errors as well as the different methods to derive the various parameters
used in the following analysis are sources of uncertainties. Those mostly affecting our analysis are the 
uncertainties on the measure of the extinction corrected H$\alpha$ and UV fluxes, thus on the H$\alpha$ to FUV flux ratio, 
the variable analyzed in this work.\\

{\bf H$\alpha$ flux:} the typical error in the observed H$\alpha$+[NII] flux is $\sim$ 0.15 mag; this includes both
the uncertainty on the spectro-photometric zero point and on the sky background subtraction 
for the extraction of the total galaxy flux in extended sources. The correction for the [NII] contamination
has a mean uncertainty of $\sim$ 0.15 mag whenever integrated spectra are available, but up to $\sim$ 0.20 mag
when a statistical correction is applied. Extinction corrections using the observed Balmer 
decrement introduce an extra 0.15 mag uncertainty that increases to $\sim$ 0.60 mag
in the case of a statistical corrections. While the S/N increases with luminosity in both
spectroscopic and narrow band imaging data, the underlying Balmer absorption, dust extinction
and [NII] contamination are more severe in bright, massive galaxies.
We thus expect that at a first order the uncertainty on the measured H$\alpha$ flux of the target galaxies
is $\sim$ 0.26 mag (0.10 dex) whenever spectroscopic data are available, and  $\sim$ 0.65 mag (0.26 dex) elsewhere.
Other sources of uncertainties hardly quantifiable are however present in the determination of 
the ionizing flux emitted by stars: these are the escape fraction and the fraction of ionizing photons absorbed
by dust before they ionize the gas ($1-f$).\\
{\bf UV flux:} the typical photometric error (zero point and
flux extraction in extended sources) in the determination of the observed FUV and NUV fluxes
is of the order of $\sim$ 15\%, with slightly better values for galaxies observed in the MIS/NGS
with respect to those observed during the AIS. The major source of uncertainty in the determination 
of the emitted UV fluxes is the extinction correction. The error on the correction is of the order of $\sim$ 0.2 mag when IR
data are available (this including $\sim$ 15 \% error on the IRAS fluxes), and 0.5 mag when statistical 
corrections are applied (Cortese et al 2008). Considering an extra 0.20 mag uncertainty due to the adopted dust model (geometry, extinction curve...)
the resulting error on the UV fluxes ranges from $\sim$ 0.32 mag (0.13 dex) in galaxies
with available far-IR data up to $\sim$ 0.56 mag (0.22 dex) whenever more indirect recipes are applied.\\
{\bf Stellar masses:} the photometric error on the near-IR magnitudes used on the 
determination of stellar masses is of the order of 0.10-0.15 mag, slightly larger on colours.
The comparison of the mass determination using different spectro-photometric models of galaxy evolution 
have shown that differences in the determined stellar masses of evolved, late-type galaxies
such as those analyzed in this work might be up to $\sim$  0.30 dex (Conroy et al. 2009a). \\
{\bf Metallicity:} the estimated value of the metallicity of the observed galaxies is rather
uncertain not only because of observational errors (flux extraction, extinction correction)
but also because of the use of indirect calibrations. It is rather difficult to quantify this 
error (Kewley \& Ellison 2008), that should be of the order of 0.20 in units of 12+Log(O/H) excluding 
the rather large uncertainties on the yields.\\
{\bf Other quantities:} the error on the $(FUV-H)_{AB}$ colour and on the effective surface brightness $\mu_{e}$(H)$_{AB}$ is $\sim$ 0.20 mag,
while that on the gas column density is small, 0.12 dex, whenever HI maps are available, while large, up to 0.33 dex, when 
gas column densities are determined using data of unresolved objects. The error on the gas fraction is of the order of $\sim$ 0.16 dex.
The resulting error on the star formation activity of the target galaxies is of a factor of 2-3 when low quality data are used, 
50 \% using high quality data.

\section{D: Possible selection biases}

The selected sample combined with the available incomplete set of data
might introduce unwanted biases in out analysis.
The [NII] emission contaminates the H$\alpha$ emission mostly near the galaxy nucleus (Boselli
2007). The correction applied to the data does not take into account this effect. The use 
of integrated spectroscopic values combined with narrow band H$\alpha$+[NII] data should
however minimize any major aperture effect. Furthermore we exclude in the following analysis
AGNs, i.e. those galaxies where the [NII] contamination is more important.\\
A possible bias in the determination of the metallicity can result from the fact that,
although we use the mean value of four different calibrations, the lack of the
[OII] 3727 \AA ~ line (and sometimes of the [OIII] 5007 \AA) in the massive galaxies 
forces us to use different calibrations as a function of the galaxy luminosity.\\
Because of the impossibility of a direct measure of the H$\alpha$ underlying absorption 
on the spectra caused by the blending of the [NII] doublet,
H$\alpha$ fluxes in the spectroscopic data have been corrected using the canonical value of 1.5 \AA. 
Detailed Balmer-line models by Kurucz (1992) indicate that the underlying absorption of the 
H$\alpha$ and H$\beta$ lines are similar within 30\%. The underlying
absorption under H$\beta$ measured from our own data is slightly larger ($\sim$ 5 \AA, Gavazzi et al. 2004),
thus the applied correction might be underestimated. A possible bias, if present, is however expected to be small
since given the line ratio of the H$\alpha$ and H$\beta$ lines (2.86 without extinction),
the error on the H$\alpha$ line is small in those objects where H$\beta$ can be detected. 
Furthermore it has been shown that the Balmer underlying absorption does not 
depend sensitively on metallicity (Kurucz 1992; González-Delgado \& Leitherer 1999),
thus this bias, if present, would hardly introduce any systematic trend with mass and metallicity.

\end{document}